\documentclass[10pt,a4paper]{article}
\usepackage{graphicx} 
\usepackage{caption}
\usepackage[margin=10pt,font=small,labelfont=bf,labelsep=endash]{caption}
\usepackage{color}
\usepackage{cite}
\usepackage{amsmath}
\usepackage{amsfonts}
\usepackage{amssymb}
\usepackage{mathtools}
\usepackage[perpage]{footmisc}
\begin{document}

\title{Liquid scintillators neutron response function: a tutorial
}

\author{M Cecconello\\
\noindent{\it\normalsize Department of Physics and Astronomy, Uppsala University, SE-751 05 Uppsala, Sweden}}

\maketitle

\begin{abstract}
This tutorial is devoted to the understanding of the different components that are present in the neutron light output pulse height distribution of liquid scintillators in fusion relevant energy ranges. The basic mechanisms for the generation of the scintillation light are briefly discussed. The different elastic collision processed between the incident neutrons and the hydrogen and carbon atoms are described in terms of probability density functions and the overall response function as their convolution. The results from this analytical approach is then compared with those obtained from simplified and full Monte Carlo simulations. Edge effect, finite energy resolution, light output and transport and competing physical processes between neutron and carbon and hydrogen atoms and their impact on the response functions are discussed. Although the analytical treatment here presented allows only for a qualitative comparison with full Monte Carlo simulations it enables an understanding of the main features present in the response function and therefore provides the ground for the interpretation of more complex response functions such those measured in fusion plasmas. Although the main part of this tutorial is focussed on the response function to mono-energetic 2.45 MeV neutrons a brief discussion is presented in case of broad neutron energy spectra and how these can be used to infer the underlying properties of fusion plasmas via the application of a forward modelling method.\\



\noindent{\bf E-mail: \tt marco.cecconello@physics.uu.se}

\end{abstract}

\section{Introduction}
\label{intro}
The measurement of the neutron emission from deuterium-deuterium (DD) and deuterium-tritium (DT) fusion reactions is one of the most important methods of assessing the performance of present and future fusion reactors. Since a neutron is released for each fusion reaction occurring in the plasma, the measurement of neutron flux emitted from the plasma is directly correlated to the fusion power. The emitted neutrons' energy spectrum is characterized by two main components at 2.45 and 14.1 MeV from DD and DT reactions respectively\footnote{In fusion devices operated with DT fuel, due to their different cross-sections, the DT neutron emission dominates over the DD one while in D-only fuel devices DT reactions can be observed at a very low level where T is generated via one branch of the DD reaction.}. 
The neutron energy spectrum is affected, among other things\footnote{Other parameters that affect the neutron production are the plasma density and the plasma effective charge but are not discussed here as they mainly affect the neutron yield and not the energy spectrum.}
, by the fusion plasma operating conditions. For example, the broadening of the Gaussian energy spectrum for 2.45 and 14.1 MeV neutrons is proportional to the square root of the plasma fuel ion temperature due to the relative velocity distribution of the reactants \cite{Jarvis}. In addition, the neutron energy spectrum is affected by the different additional heating schemes that are normally employed in fusion devices such as neutral beam injection and radio-frequency heating. 
For example, in the case of neutral beam heating the emitted neutron energy spectrum is the combination of a thermal component due to the plasma fuel ions, a beam-thermal component from the reactions between beam and fuels ions and a beam-beam component. The relative intensity of these components is in turn affected by the plasma conditions. In present day devices, the beam-thermal component is dominating while in future fusion devices, the thermal component will be dominant. 
In the case of radio-frequency heating, neutrons with energies above the DD and DT energies are observed due to fusion reaction from ions accelerated to energies of a few MeV. 
The measurement of the spatial and temporal evolution of the neutron emission from fusion plasmas in terms of its flux and energy spectrum can therefore provide important information on the plasma itself which can be used to optimize fusion power production.

Different diagnostics are used to measure the neutron emission from fusion plasmas  \cite{Jarvis, Wolle} but all rely on the conversion of the neutron into a charged particle that can then be detected. 
This conversion takes places either thanks to neutron induced nuclear reactions in which heavy charged fission fragments are produced\footnote{Thermal neutron capture induced reaction in $^{235}$U is a typical example.} or via elastic scattering of neutrons with light atoms, typically hydrogen. 
In their simplest form,  neutron diagnostics can just be used as counters where the measured counts are proportional to the number of neutrons emitted by the plasma and therefore to the fusion power. 
The proportionality constant is determined via the absolute calibration of such neutron counters and depends on several parameters such as the counter's efficiency and its position with respect to the neutron source (the plasma) and the fusion reactor \cite{Syme, Batistoni}. 
Neutron spectrometers are more sophisticated diagnostics in which the measured neutron energy spectrum is linked in a non-trivial way to the neutron source. Since the energy of fission fragments does not reflect the energy spectrum of the incident neutron, fission chambers can not be used as spectrometers, if one excludes the very crude spectroscopic capability offered by threshold reactions. 
For this reason, neutron spectrometers are all based on the conversion of the neutron into a light recoil particle via elastic scattering. Neutron spectrometers can be distinguished by the way in which the scattered neutron and the recoil particle are processed. 
In compact spectrometers, the recoil particle deposits its energy into the scattering medium which, depending on the material, can emit a pulse of scintillation light that is detected: in this case the scatterer itself acts as the detector. The light emission is then converted into a voltage signal and the voltage pulse height spectrum generated by the recoil particles is measured\footnote{Pulse height spectra can be based on the peak amplitude or on the time integrated voltage signal.}. Since the voltage signal is proportional to the energy deposited by the recoil particle and this depends on the incident neutron energy, a detector based on scintillation material can, in principle, be used as a neutron spectrometer. 
Large spectrometers, instead, are based on the measurement of the scattered neutron or recoil particle in a detector other than the scatterer. For example, the recoil protons ejected by neutron scattering on a thin, hydrogen-rich foil can be detected in an array of detectors after they have been momentum and energy separated \cite{Eric}. Alternatively,
the time difference between the two scintillation events generated by the same incident neutron in two spatially separated scintillators provides the neutron time of flight and therefore its energy \cite{Maria}. Recently, time-of-flight spectrometers are also taking advantage of the information on the amount of energy deposited by the recoil particles in the scatterer and in the detector to suppress the contribution from unwanted random coincidences that, especially in fusion devices, typically affect such instruments \cite{Skiba}.

The often complex relation between the incident neutron energy spectrum and the output signal from the detector is referred to as the detector response function. Since the  neutron energy spectrum at the detector's location is not mono-energetic and since the response function is usually dependent on the incident neutron energy, it is necessary to determine the response function for all the neutron energies of interest. The resulting set of response functions (one for each incident neutron energy) is referred to as the detector response function matrix.\footnote{The response function for large spectrometers, which takes into account for example the effect of the time of flight geometry between scatterer and detector, is referred to as the instrument response function to distinguish it from the individual detectors' response functions.}
It is the knowledge of this response function that allows to infer the characteristics of the neutron energy spectrum and ultimately of the plasma itself. Two different approaches can be used to relate the measured neutron energy spectrum to the fusion plasma source: \emph{i}) forward modelling and \emph{ii}) inversion algorithms. 
Forward modelling relies on the accurate modelling of all the processes that in the plasma affect the neutron emission, of the neutron transport from the source to the detector, of the conversion of the incident neutron field into recoil particles and eventually into the detector output signal \cite{Binda}. Inversion algorithms make no assumption or modelling on the neutron source which instead is obtained by different least-squares minimization methods based on the knowledge of the detector response function \cite{Zimbal}. Both methods have advantages and disadvantages but the one thing they have in common is the requirement for a very well characterized detector response function. 
Response functions for scintillator detectors are usually measured experimentally with well characterized neutron sources \cite{Klein, Enqvist} and interpreted with the help of dedicated Monte Carlo codes such as NRESP \cite{NRESP} or more general radiation transport codes such as MCNP \cite{MCNP} and MCNP-PoliMi \cite{Pozzi} and GEANT4 \cite{Agostinelli} (which has the additional feature of simulating the scintillation light transport too). Very good agreement is found between measured and simulated neutron response functions for all the cited codes. For the purpose of this tutorial, NRESP will be used. 
In NRESP, the relevant cross-sections (and differential cross-sections) for neutron interactions with hydrogen and carbon in the energy range 0.02 to 20 MeV are included. The detector material composition and geometry including the liquid scintillator housing and in the optical window connecting to the photo-cathode are modelled. The deposited energy is then converted into a light pulse height distribution taking into account the finite detector energy resolution and experimentally measured light output functions for all the generated particles in the medium (recoils and $\alpha$-s from nuclear reactions on carbon). 

The interpretation of the measured or simulated response functions for liquid scintillator can be quite difficult if only the light output pulse height distribution is given. For a detailed understanding of the origin of the different features present in the response function it is necessary to analyse the contributions from the individual processes occurring in the liquid scintillator such as single and multiple elastic scattering, nuclear reactions and so on. This information can be obtained from the Monte Carlo codes discussed above but this is usually not trivial. In addition, although these codes can provide the expected light output pulse height spectrum, for example, for neutrons that have collided three times with protons, they do no provide an explanation for why it has that particular shape. The aim of this tutorial is to provide such an explanation by means of an analytical derivation of the expected light output pulse height distribution combined with a very simple Monte Carlo code used for its verification. The analytical approach is limited to a few simple cases but it provides nevertheless the basic understanding of how the real response function arises from a combination of multiple individual neutron interactions.
This tutorial focuses in particular on the detailed explanation of the different contributions to the response function of a liquid scintillator to incident mono-energetic neutrons. As it will be shown, the interpretation of the response function even in this simple situation is far from trivial. The results obtained in this particular case are easily generalized for a broad neutron energy spectrum.
The tutorial is therefore structured as follows. Section \ref{sec:EJ301} is dedicated to a brief overview of the scintillation process, of the resulting recoil particle light output response function and of a simplified Monte Carlo code used for the study of multiple elastic scattering. Section \ref{sec:singleNP} is devoted to the study of the light response function in the case of single elastic scattering processes. Section \ref{sec:doubleNP} is dedicated to the determination of the light response function in the case of multiple elastic scattering of a neutron with particles of the same species (for example, only with protons or only with carbon atoms). Section \ref{sec:mixedNP} discusses the mixed double scattering case in which the neutron scatters elastically with two different particles (for example first with a carbon atom and then with a hydrogen atom). In section \ref{sec:Comparison}, the light response functions calculated with full Monte Carlo simulations are qualitatively interpreted with the help of the response functions derived  in the previous two sections for each elastic scattering process. Section \ref{sec:complications} briefly addresses some aspects affecting the light response function that have been neglected in the previous sections and elucidate the use of the response function matrix for the interpretation of the liquid scintillator response function when non mono-energetic neutron sources are present such as in the case of fusion reactions. In addition, the forward modelling method is also discussed. Final comments and remarks are presented in section \ref{sec:summary}.

\section{Neutron response function of liquid scintillators}
\label{sec:EJ301}
Scintillators are among the most common type of detectors both for $\gamma$-rays and neutron radiation detection and operate on the principle of induced fluorescent light emission upon the interaction of the radiation within the material. 
A detailed description of the physical principles, material composition and application of scintillators can be found in \cite{Knoll}. For some scintillation materials, light emission depends on the type of incident radiation and therefore it is possible to discriminate between $\gamma$-rays and neutrons, a feature that is essential in fusion application as neutron diagnostics often operates in mixed fields. 
Liquid scintillators based on organic materials such as benzene (C$_6$H$_6$), tuolene (C$_6$H$_5\cdot$CH$_3$) and xylene (C$_6$H$_4\cdot$(CH$_3$)$_2$), that is hydrogen-rich materials, belong to this category. The main interaction mechanism between $\gamma$-rays and liquid scintillators is Compton scattering with the electrons while neutrons interacts by elastic scattering with the hydrogen and carbon nuclei\footnote{Inelastic scattering and nuclear reactions with carbon atom are also possible for neutron energies above 4 MeV and are discussed briefly later.}. 
In both cases a recoil particle is generated (an electron for $\gamma$-rays and a proton or carbon nucleus for neutrons). Coulomb interactions between recoil particles and the organic scintillator molecules result in the conversion of the recoil kinetic energy into molecular excitation energy. Part of this excitation energy is then dissipated via thermal quenching and part via fluorescent light emission in the UV region of the visible spectrum. Typical light pulses have a very fast rise time (few nanoseconds) and decay constants between 20 and 200 ns.
\begin{figure}
	\begin{center}
  \includegraphics[scale = 0.6]{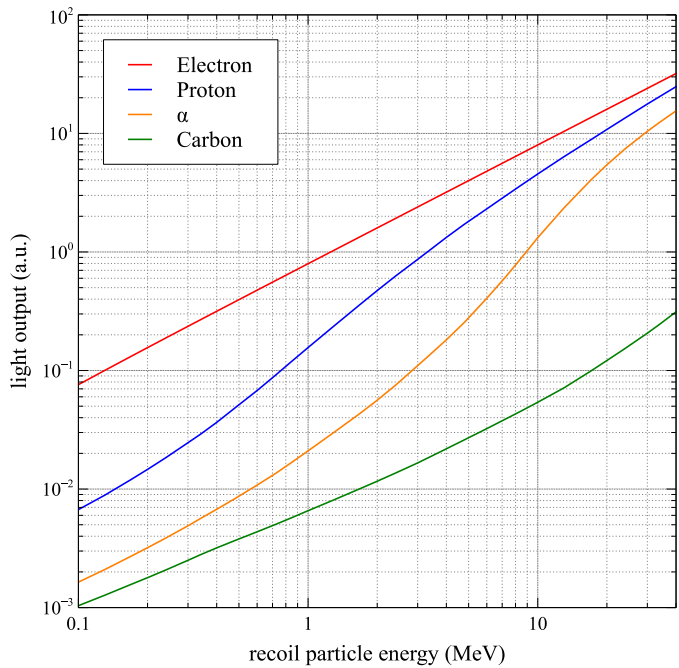}
	\end{center}
\caption{Liquid scintillator light output as a function the recoil particle energy. Data adapted from \cite{Verbinski}.}
\label{fig:Verbinski}       
\end{figure}
The fluorescent light is then transported (either directly of via reflection with the liquid scintillator housing walls) to a photocathode: depending on the volume of the scintillator, significant attenuation of the fluorescent light in the scintillator medium itself can occur.
The fluorescent light is then converted into electrons at the photocathode via the photoelectric effect and the initial few ejected photo-electrons are subsequently accelerated, focussed and multiplied via secondary electron emission by multi-stage dynodes photomultiplier resulting in large gains ($10^5$ - $10^9$). The conversion of photons into electrons and their multiplication is a process that, under the correct experimental conditions, is highly linear. Non-linearities occur for example if the photomultiplier is operated at high gains, at high counting rates and in presence of even weak magnetic fields.
The electron current at the anode of the photomultiplier is converted into a voltage via a load resistor and the detector voltage output is fed into the acquisition system by co-axial cables (usually tens of meters in fusion experiments). Attenuation and distortion of the voltage signal in the cables will occur but these processes are linear and easily modelled.
In present days fusion neutron diagnostics the detector voltage signal is digitized at very high sampling frequencies ranging from 250 MHz up to 4 GHz with high resolution from 10 to 14 bits. The ADC process can be considered linear if the integral an differential non-linearities are negligible which is often the case.
Several conversion mechanism contribute therefore to the final signal that is measured for a single neutron interacting with the detector. If the liquid scintillator is to be used as a spectrometer it is then important that a good linearity exists between the recorded signal and in the incident neutron energy.
Under the correct experimental conditions all the above processes are linear but one: the fluorescent light output for recoil particles from neutron elastic collision is inherently non-linear as shown in figure \ref{fig:Verbinski}. This non-linearity complicates significantly the neutron response function as discussed in detail in sections \ref{sec:singleNP}, \ref{sec:doubleNP} and \ref{sec:mixedNP}.
\begin{figure}
		\begin{center}
  \includegraphics[scale = 0.65]{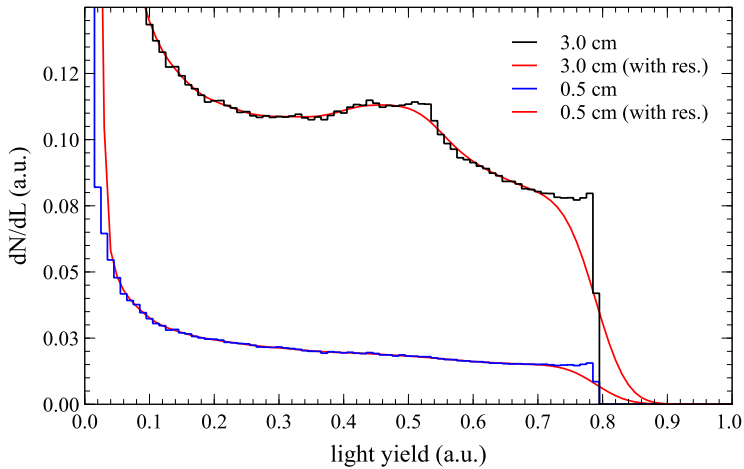}
  	\end{center}
\caption{Example of the light response response function of a liquid scintillator to a mono-energetic neutron with energy of 2.45 MeV for a liquid scintillator of 6 cm diameter and two different thicknesses with and without the detector energy resolution effect included.}
\label{fig:RFMEN}       
\end{figure}

Examples of the response function calculated with NRESP of two liquid scintillators with the same diameter (6 cm) and two different thickness (0.5 and 3.0 cm) for a beam of incident mono-energetic neutrons with an energy of 2.45 MeV are shown in figure \ref{fig:RFMEN}.
As can be seen, the response function exhibits a rich series of features some of which depends on the detector thickness. For the purpose of this tutorial a Simplified monTe cArlo neutron Response funcTion Simulator (STARTS) has been written to calculate the probability density function of the energy and light output distributions of multi-scattered neutrons and of all recoil protons and carbon nuclei in any possible combination. Contrary to the full Monte Carlo codes discussed above, STARTS is specifically aimed at the calculation of the pulse height spectra for specific types of elastic collisions.
The following simplifications have been made in STARTS: \emph{i}) no cross-section dependence is included as the particles involved in the elastic scattering are selected by the user, \emph{ii}) all neutrons undergo a fixed number of elastic scattering defined by the user, \emph{iii}) the detector is considered infinite in size, \emph{iv}) recoil particles deposit all their energy in the detector, \emph{v}) the incident neutrons are mono-energetic and \emph{vi}) no energy resolution broadening is included in the calculation of the response function and \emph{vii}) the light output function used is taken from \cite{NRESP}.
The first simplification implies that the relative intensity of the contribution to the response function from neutron-proton (np) and from neutron-carbon (nC) elastic collisions  is neglected: this does not affect the shape of the response function. Simplifications \emph{iii}) and \emph{iv}) do not affect in any qualitative way the response function while simplification \emph{vi}) has an important effect which is however well known and easily understood. The impact of all these simplifications is briefly discussed in section \ref{sec:singleNP}.

\section{Response function for single neutron scattering}
\label{sec:singleNP}
\begin{figure*}
	\begin{center}
  \includegraphics[width=0.9\textwidth]{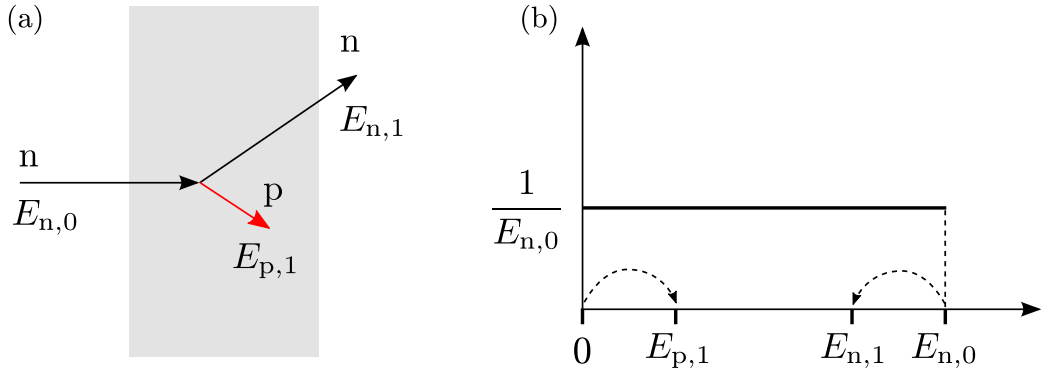}
  	\end{center}
\caption{Panel (a): depiction of an incoming neutron undergoing single elastic scattering resulting in a scattered neutron and a recoil proton; the shaded area indicates the liquid scintillator volume. Panel (b): probability density function for the energy of the scattered neutron.}
\label{fig:SingleScattering}       
\end{figure*}

Consider an incident neutron with initial energy $E_\mathrm{n,0}$ making an elastic collision with a proton, assumed to be at rest, as depicted schematically in panel (a) of figure \ref{fig:SingleScattering}. After the elastic collision the energies of the neutron and of the recoil protons will be $E_\mathrm{n,1}$ and $E_\mathrm{p,1}$ where the index ``1'' indicates that these quantities refer to the energy of the particles after the first collision. From classical mechanics, invoking the conservation of energy and linear momentum, it can be shown that in the case of a neutron colliding with a generic target particle these energies are given by:
\begin{align}
\label{eq:En1}
	& \dfrac{E_\mathrm{n,1}(\theta)}{E_\mathrm{n,0}} = \dfrac{(1+\alpha) + (1-\alpha) \cos \theta}{2} \\ 
\label{eq:Et1}	
	& \dfrac{E_\mathrm{t,1}(\theta)}{E_\mathrm{n,0}} =\dfrac{(1 - \alpha)(1-\cos\theta)}{2} 
\end{align}
where the index $t$ identifies the recoil target, $\theta$ is the scattering angle in the centre of mass reference system, $\alpha = (A-1)^2/(A+1)^2$ and $A = m_\mathrm{t}/m_\mathrm{n}$ that is the ratio between the target and neutron masses. It is useful, especially for the discussion in the following sections, to observe that $E_\mathrm{n,1}/\alpha$ represent the maximum possible energy from which a neutron with energy $E_\mathrm{n,1}$ could have originated in a single elastic scattering.
Under the assumption of an isotropic cross-section for the elastic scattering in the centre of mass, the scattering angle can assume any value between 0 and $\pi$ with equal probability\footnote{This assumption is a good approximation for np elastic collision for neutrons of few MeV and slightly less accurate for the nC elastic collision case in which forward scattering is more favourable. For the sake of simplicity, this effect is here neglected as it makes a small difference to the calculations and results here discussed.}.
It can then be shown that the probability of the scattered neutron to have an energy in the interval $[E_\mathrm{n,1}, E_\mathrm{n,1} + dE_\mathrm{n,1}]$ is given by:
\begin{equation}
\label{eq:pdfSCn}
p(E_{\mathrm{n},1}) dE_{\mathrm{n},1} = 
\begin{cases}
	\dfrac{1}{1-\alpha} \dfrac{1}{E_\mathrm{n,0}} dE_{\mathrm{n},1} & \hspace{0.1cm} \mathrm{if~} E_{\mathrm{n},1} \in [\alpha E_\mathrm{n,0}, E_\mathrm{n,0}] \\[14pt]
	0  & \hspace{0.1cm} \mathrm{otherwise} 
\end{cases}
\end{equation}
where $p(E_{\mathrm{n},1})$ is the Probability Density Function (PDF) of the continuous random variable $E_{\mathrm{n},1}$. Note that $p(E_{\mathrm{n},1})$ is a properly normalized PDF\footnotemark. In fact, the total probability of the neutron having an energy in the interval $[\alpha E_\mathrm{n,0}, E_\mathrm{n,0}]$ after one collision is the sum (integral) of all the probabilities of the neutron having an energy in the interval $[E_\mathrm{n,1}, E_\mathrm{n,1} + dE_\mathrm{n,1}]$ over all possible energies, which is equal to:
\begin{equation}
\label{eq:TotPSCn}
	\int_{\alpha E_\mathrm{n,0}}^{E_\mathrm{n,0}} p(E_{\mathrm{n},1}) dE_{\mathrm{n},1} = 1
\end{equation}
as it can be easily verified by inserting equation \eqref{eq:pdfSCn} into the above expression. note that equation \eqref{eq:TotPSCn} is the total probability law for mutually exclusive events for continuous random variables.
\footnotetext{\label{fn:probdef} This probability can also be thought as the probability of the neutron having that energy given that an elastic collision has occurred. The latter probability is linked to the detection efficiency as discussed in section \ref{sec:efficiency}}.

Similarly, the probability of the recoil target of having an energy in the interval $[E_\mathrm{t,1}, E_\mathrm{t,1} + d E_\mathrm{t,1}]$ is given by:
\begin{equation}
\label{eq:pdfSCt}
p(E_{\mathrm{t},1}) dE_{\mathrm{t},1}=
\begin{cases}
	\dfrac{1}{1-\alpha} \dfrac{1}{E_\mathrm{n,0}} dE_{\mathrm{t},1} & \hspace{0.1cm} \mathrm{if~} E_{\mathrm{t},1} \in [0, (1 - \alpha) E_\mathrm{n,0}] \\[14pt]
	0  & \hspace{0.1cm} \mathrm{otherwise} 
\end{cases}.
\end{equation}

According to equations \eqref{eq:pdfSCn} and \eqref{eq:pdfSCt}, the probability of observing a scattered neutron and recoil target with energies in the range $[E, E+dE]$ is non-zero and constant within the appropriate energy interval as shown in panel (b) of figure \ref{fig:SingleScattering}. In the case of np elastic scattering, $\alpha = 0$ which implies that the recoil proton energy $E_\mathrm{p,1}$ can assume any value between 0 MeV (``grazing'' collision) and $E_\mathrm{n,0}$ (``head-on'' collision) with equal probability. In the particular case depicted in panel (b) of figure \ref{fig:SingleScattering}, the energy of the scattered neutron is $E_\mathrm{n,1}$ and the energy of the recoil proton is $E_\mathrm{p,1} = E_\mathrm{n,0} - E_\mathrm{n,1}$.  

Since for mono-energetic neutrons with energy $E_\mathrm{n,0}$ the recoil protons can equally assume energies in $[0, E_\mathrm{n,0}]$ it follows that, in this simplified scenario, the response function of the detector is the one depicted in panel (b) of figure \ref{fig:SingleScattering}, that is a ``box-like'' function. 
As can be seen from figure \ref{fig:Verbinski}, a recoil proton of energy $E_\mathrm{p,1}= E_\mathrm{n,0} = 2.45$ MeV would generate a light output yield of approximately $L \approx 0.8$ which is exactly the upper edge of the response functions shown in figure \ref{fig:RFMEN}.

\begin{figure}
		\begin{center}
  \includegraphics[scale = 0.55]{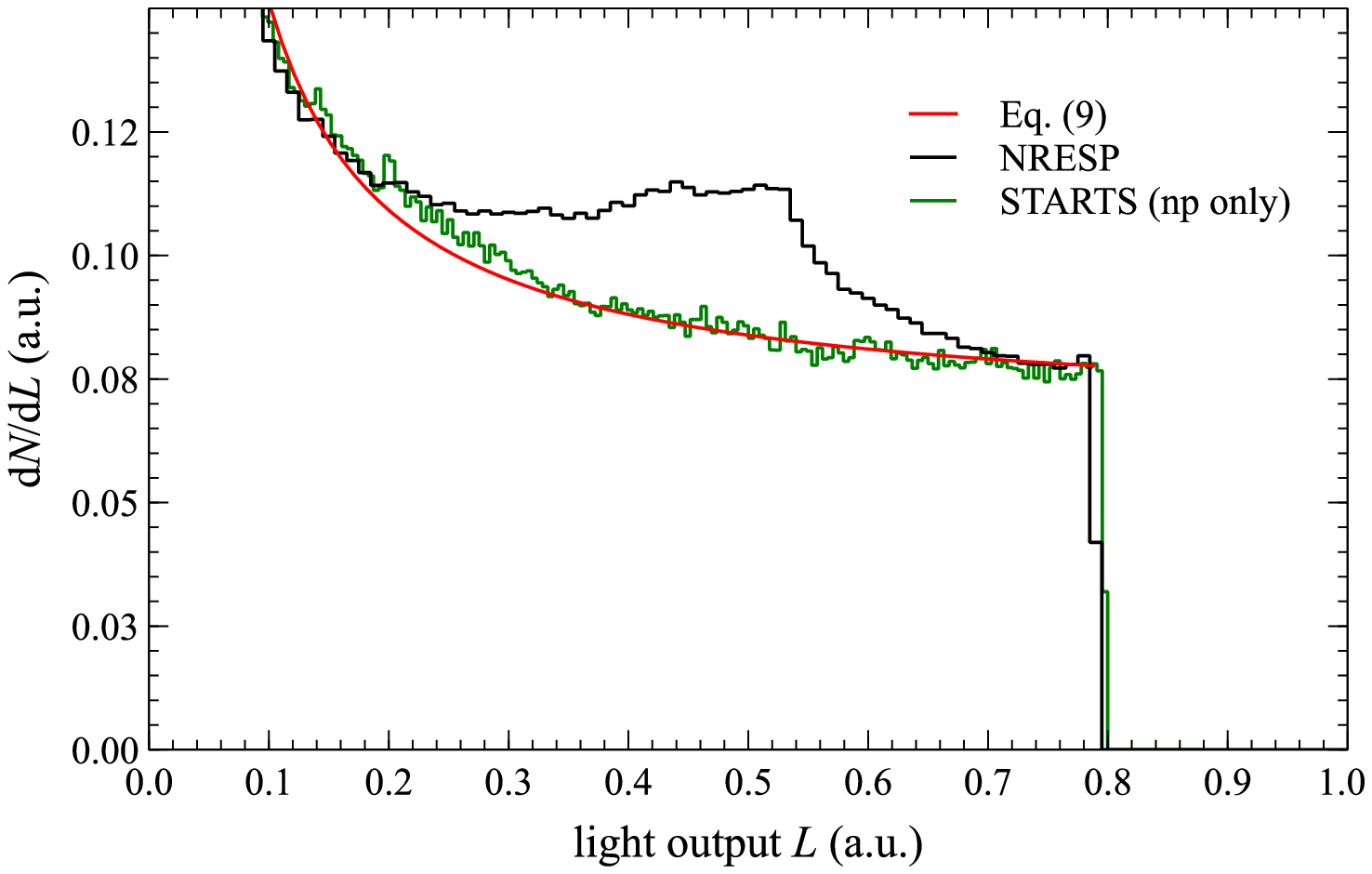}
  	\end{center}
\caption{Example of the light response response function of a liquid scintillator to a mono-energetic neutron with energy of 2.45 MeV for a liquid scintillator of 6 cm diameter and two different thicknesses with and without the detector energy resolution effect included.}
\label{fig:LRFEp}       
\end{figure}

This simple ``box-like'' response function is modified by the non-linear dependence of the light output function on the recoil particle's energy. Following \cite{Knoll}, if one assumes for the light output function the relation:
\begin{equation}
\label{eq:LO}
	L(E_\mathrm{p}) = k E_\mathrm{p}^\beta
\end{equation}
then:
\begin{equation}
\label{eq:invLO}
	E_\mathrm{p}(L) = \left( \dfrac{L}{k} \right)^{1/\beta}
\end{equation}
and therefore:
\begin{equation}
\label{eq:dEpdL}
	\dfrac{\mathrm{d} E_\mathrm{p}(L)}{\mathrm{d} L} = \dfrac{1}{\beta L} \left( \dfrac{L}{k} \right)^{1/\beta}
\end{equation}
where $k \approx 0.21$ MeV$^{-\beta}$ and $\beta \approx 3/2$ give a good approximation to the light output function for recoil protons shown in figure \ref{fig:Verbinski} up to 3 MeV. 
Equations \eqref{eq:pdfSCt} (with $\alpha = 0$) and \eqref{eq:dEpdL} can be combined to obtain the light response function for the recoil protons:
\begin{equation}
\label{eq:NLOSnp}
\dfrac{\mathrm{d} N}{\mathrm{d} L} = \dfrac{\mathrm{d} N}{\mathrm{d} E_\mathrm{p,1}} \dfrac{\mathrm{d} E_\mathrm{p,1}}{\mathrm{d} L} = \dfrac{1}{E_\mathrm{n,0}} \dfrac{1}{\beta L}  \left( \dfrac{L}{k} \right)^{1/\beta}
\end{equation}
which implies that the light output response function increases for low light yields. Figure \ref{fig:LRFEp} shows the light response function for the 3 cm thick detector calculated by NRESP, by equation \eqref{eq:NLOSnp} and by STARTS. As can be seen, the overall features at low light output yields ($L \lesssim 0.2$) and for $L \approx 0.8$ are well described qualitatively\footnote{The term ``qualitative'' is here use to indicate that the absolute amplitude of the pulse height spectrum can not be calculated by STARTS and ``ad-hoc'' scaling factors are used instead.} but not in the range in between. 

The ``bump'' seen in figures \ref{fig:RFMEN} and \ref{fig:LRFEp} at intermediate $L$  values can not be explained as the result of the detector finite resolution, of edge effects for detectors of finite size nor as a consequence of the contribution from nC scattering. The effect of the detector finite energy resolution is mainly to ``smear out'' the sharp edge at the maximum light output as shown by the red curves in figure \ref{fig:RFMEN}.
For a detector of finite size it is possible for the recoil protons generated near the outer surface of the scintillator to escape after having deposited in the medium only a fraction of their energy. The mean free path $\lambda$ of 2.45 MeV protons in liquid scintillators is of the order of $\lambda \lesssim 0.1$ mm.
Even assuming that all the recoil protons generated within a distance $\lambda$ from the outer surfaces escape, the overall light response function would be reduced in its amplitude by approximately 1 \% but its shape would not be modified. The effect would be even smaller for recoil protons of lower energies. According to equation \eqref{eq:Et1}, the maximum energy for a recoil carbon, for which $\alpha \approx 0.716$, is $E_\mathrm{c,1} = (1-\alpha) E_\mathrm{n,0} \approx 0.7$ MeV. The corresponding light output is $L \approx 0.01$ (see figure \ref{fig:Verbinski}) and therefore the contribution to the light response function from recoil carbon nuclei is confined to the very low end of the response function. On the scale used in figure \ref{fig:LRFEp} this is hardly visible.

It is possible to conclude therefore that the ``box-like'' response function for single np elastic scattering combined with the light output non-linearity together with the detector finite energy resolution describe quite well the simulated response function especially for the thin detector. However, for thick detectors this is not the case: in \cite{Knoll} this feature is described as the result of neutron double elastic scattering with protons in the scintillator. Section \ref{sec:doubleNP} is dedicated to the understanding of the origin and shape of this feature.
\begin{figure}
	\begin{center}
  \includegraphics[scale = 0.5]{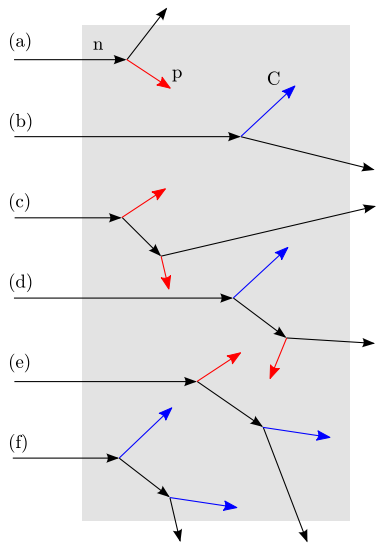}
  	\end{center}
\caption{Single and double elastic scattering events for a neutron incident on a liquid scintillator (gray shaded area). Tracks (a) and (b) are single scattering with a proton and carbon atom respectively. Tracks (c) to (d) represent the four possible collisions: np \& np (c); nC \& np (d); np \& nC (e) and nC \& nC (f).} 
\label{fig:DoubleScattering}       
\end{figure}

\section{Response function for multiple neutron scattering}
\label{sec:doubleNP}
In the case of an incident neutron making two elastic collisions four different outcomes are possible as shown in figure \ref{fig:DoubleScattering}: double elastic collision on two protons (track ``c'') or on two carbon nuclei (track ``f'') or mixed scattering first on a carbon nucleus and then on a proton (track ``d'') or the other way around (track ``e''). 
As discussed in section \ref{sec:singleNP}, the contribution to the total light output  from recoil carbon nuclei is negligible and therefore the double scattering on carbon nuclei can also be neglected and is not further discussed here. In a similar fashion, the total light output in the case in which the neutron first collides with a proton and then with a carbon nucleus is almost equivalent to a single elastic scattering with a proton. Tracks of type ``e'' therefore contributes to the response function as single np events described in section \ref{sec:singleNP}. 
The reponse function from double scattering on protons (track ``c'') is discussed in this section while track ``d'' (collision on carbon nucleus followed by collision on proton) is discussed in section \ref{sec:mixedNP}.
The discussion of the neutron double elastic scattering on proton is divided in four parts. The probability density function for the energy of the neutron after two elastic collisions is derived in section \ref{sec:doubleNP_projectile} while the probability density function for the energy of the second recoil target is derived in section \ref{sec:doubleNP_target}. The probability density function for the total deposited energy in two elastic collisions by the neutron is derived in section \ref{sec:doubleNP_DepEnergy} and finally the PDF for the corresponding total light output is obtained in section \ref{sec:doubleNP_LightOutput}.

\subsection{Doubly scattered neutron energy probability density function}
\label{sec:doubleNP_projectile}
The probability of a neutron with initial energy $E_\mathrm{n,0}$ having energy in the range $[E_\mathrm{n,2}, E_\mathrm{n,2} + dE_\mathrm{n,2}]$ after two elastic collisions given that after the first collision it had an energy in the range $[E_\mathrm{n,1}, E_\mathrm{n,1} + dE_\mathrm{n,1}]$ is given by the conditional probability:
\begin{equation}
\label{eq:CondProb}
 p(E_{\mathrm{n},2} \mid E_{\mathrm{n},1}) dE_\mathrm{n,2} = \dfrac{p(E_\mathrm{n,2} \cap E_\mathrm{n,1})}{p(E_\mathrm{n,1})} dE_\mathrm{n,2}  
\end{equation}
where $p(E_\mathrm{n,2} \cap E_\mathrm{n,1})$ is the PDF of the joint events resulting in the energies $E_\mathrm{n,1}$ and $E_\mathrm{n,2}$. Recalling equation \eqref{eq:pdfSCn}, the probability $p(E_{\mathrm{n},2} \mid E_{\mathrm{n},1}) dE_\mathrm{n,2}$ is given by:
\begin{equation}
	\label{eq:pdfSC2nd}
	p(E_\mathrm{n,2}|E_\mathrm{n,1}) dE_\mathrm{n,2}  = 
	\begin{cases}
		\dfrac{1}{1-\alpha} \dfrac{1}{E_{\mathrm{n},1}} dE_{\mathrm{n},2} & 			\hspace{0.1cm} \mathrm{if~} E_{\mathrm{n},2} \in [\alpha E_{\mathrm{n},1}, E_{\mathrm{n},1}] \\[14pt]
	0  & \hspace{0.1cm} \mathrm{otherwise}
	\end{cases}
\end{equation}
which is equal to equation \eqref{eq:pdfSCn} with $E_\mathrm{n,0}$ and $E_\mathrm{n,1}$ replaced by $E_\mathrm{n,1}$ and $E_\mathrm{n,2}$ respectively.

\begin{figure}
	\begin{center}
  \includegraphics[scale = 0.8]{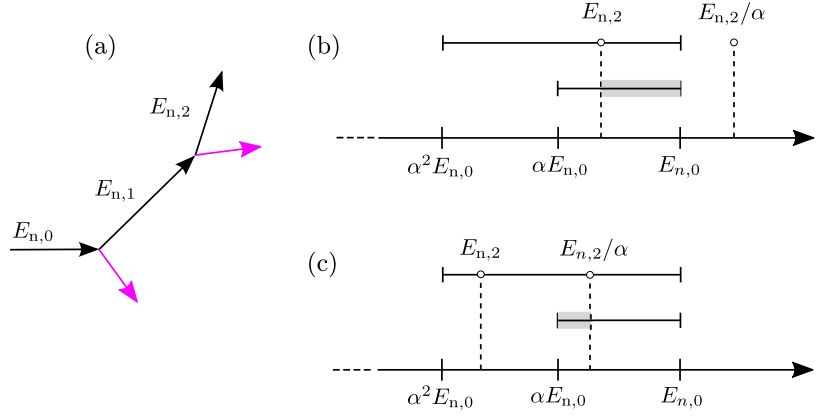}
  	\end{center}
\caption{A depiction of a doubly scattered neutron with energy $E_\mathrm{n,0}$ from the same type of target nucleus (indicated by the magenta arrows) is shown in panel (a). Panels (b) and (c) show the range of possible energies that the neutron can have after one or two  elastic scattering (short and long horizontal bars respectively). The shaded area indicates the range of possible energies for the neutron after one collision if after two collisions it has energy $E_\mathrm{n,2}$.} 
\label{fig:DSPossibleEnergies}       
\end{figure}

The probability $p(E_\mathrm{n,2})dE_\mathrm{n,2}$ of a neutron with initial energy $E_\mathrm{n,0}$ having energy in the range $[E_\mathrm{n,2}, E_\mathrm{n,2} + dE_\mathrm{n,2}]$ after two elastic collisions regardless of which energy it had after the first collision can be obtained, using the law of the total probability, as the sum of the probabilities of all the possible combination of two collisions that could have resulted in $E_\mathrm{n,2}$ being in such range. Each joint event has a probability given by $p(E_\mathrm{n,2} \cap E_\mathrm{n,1}) dE_\mathrm{n,2}$ which, using equation \eqref{eq:CondProb} and replacing $p(E_\mathrm{n,1})$ with equation \eqref{eq:pdfSCn} in combination with equation \eqref{eq:pdfSC2nd}, can be written as:
\begin{equation}
\label{eq:CondProExplicit}
	p(E_\mathrm{n,2} \cap E_\mathrm{n,1}) dE_\mathrm{n,2} = \dfrac{1}{(1-\alpha)^2}\dfrac{1}{E_\mathrm{n,0}}\dfrac{1}{E_\mathrm{n,1}} dE_{\mathrm{n,2}}.
\end{equation} 
The total probability $p(E_\mathrm{n,2})dE_\mathrm{n,2}$ is then obtained by integrating over all possible energies $E_\mathrm{n,1}$:
\begin{equation}
\label{eq:TotPbDS}
	p(E_\mathrm{n,2}) dE_{\mathrm{n,2}} = dE_{\mathrm{n,2}} \int \dfrac{1}{(1-\alpha)^2}\dfrac{1}{E_\mathrm{n,0}}\dfrac{1}{E_\mathrm{n,1}}  dE_\mathrm{n,1}.
\end{equation}
Note that $E_\mathrm{n,2}$ ranges between the neutron initial energy $E_\mathrm{n,0}$, corresponding to two subsequent ``grazing'' collisions, and the minimum energy $\alpha^2 E_\mathrm{n,0}$ corresponding to two subsequent ``head-on'' collisions ($\theta = \pi$). If $E_\mathrm{n,2} \in [\alpha E_\mathrm{n,0}, E_\mathrm{n,0}]$, then $E_\mathrm{n,1}$ can take any value between $E_\mathrm{n,2}$ and $E_\mathrm{n,0}$ (see panel (b) of figure \ref{fig:DSPossibleEnergies}) so that equation \eqref{eq:TotPbDS} results in:
\begin{align}
\label{eq:TotPbDS1}
	p(E_\mathrm{n,2}) dE_\mathrm{n,2} & = \int_{E_\mathrm{n,2}}^{E_\mathrm{n,0}} \dfrac{1}{(1-\alpha)^2}\dfrac{1}{E_\mathrm{n,0}}\dfrac{1}{E_\mathrm{n,1}} dE_\mathrm{n,1}  dE_\mathrm{n,2}\\
\label{eq:TotPbDS1integrated}
	& = \dfrac{1}{(1-\alpha)^2} \dfrac{1}{E_\mathrm{n,0}} \ln \left( \dfrac{E_\mathrm{n,0}}{E_\mathrm{n,2}} \right) dE_\mathrm{n,2}.
\end{align}
If $E_\mathrm{n,2} \in [\alpha^2 E_\mathrm{n,0}, \alpha E_\mathrm{n,0}]$, then $E_\mathrm{n,1}$ can range only between $\alpha E_\mathrm{n,0}$ and $E_\mathrm{n,2}/ \alpha$ since in a single collision it is not possible for a neutron with initial energy $E_\mathrm{n,1} \in [E\mathrm{n,2}/\alpha, E_\mathrm{n,0}]$  to reach a final energy in the range $[\alpha^2 E_\mathrm{n,0}, E_\mathrm{n,2}]$ (see panel (c) of figure \ref{fig:DSPossibleEnergies}). In this case then, equation \eqref{eq:TotPbDS} gives:
\begin{align}
\label{eq:TotPbDS2}
	p(E_\mathrm{n,2}) dE_\mathrm{n,2} & = \int_{\alpha E_\mathrm{n,0}}^{E_\mathrm{n,2}/\alpha} \dfrac{1}{(1-\alpha)^2}\dfrac{1}{E_\mathrm{n,0}}\dfrac{1}{E_\mathrm{n,1}} dE_\mathrm{n,1} dE_\mathrm{n,2}\\
\label{eq:TotPbDS2integrated}
	& = \dfrac{1}{(1-\alpha)^2} \dfrac{1}{E_\mathrm{n,0}} \ln \left( \dfrac{E_\mathrm{n,2}}{\alpha^2 E_\mathrm{n,0}} \right) dE_\mathrm{n,2}.
\end{align}
To summarize, the probability $p(E_\mathrm{n,2})dE_\mathrm{n,2}$ of a neutron with initial energy $E_\mathrm{n,0}$ having energy in the range $[E_\mathrm{n,2}, E_\mathrm{n,2} + dE_\mathrm{n,2}]$ after two elastic collisions is given by:
\begin{equation}
\label{eq:TotPDSsummary}
p(E_\mathrm{n,2}) dE_\mathrm{n,2} = 
\begin{cases}
	\dfrac{1}{(1-\alpha)^2} \dfrac{1}{E_\mathrm{n,0}} \ln \left( \dfrac{E_\mathrm{n,0}}{E_\mathrm{n,2}} \right)dE_\mathrm{n,2} & \mathrm{if~} E_\mathrm{n,2} \in [\alpha E_\mathrm{n,0}, E_\mathrm{n,0}] \\[14pt]
	\dfrac{1}{(1-\alpha)^2} \dfrac{1}{E_\mathrm{n,0}} \ln \left( \dfrac{E_\mathrm{n,2}}{\alpha^2 E_\mathrm{n,0}} \right)dE_\mathrm{n,2} & \mathrm{if~} E_\mathrm{n,2} \in [\alpha^2 E_\mathrm{n,0}, \alpha E_\mathrm{n,0}] \\[14pt]
	0  & \mathrm{otherwise} 
\end{cases}
\end{equation}
and the corresponding PDF is shown in panel (a) of figure \ref{fig:DoubleScattering}. In the particular case of the two targets being protons, $\alpha = 0$ and therefore $E_\mathrm{n,1}$ and $E_\mathrm{n,2}$ can  both take any value in the interval  $[0, E_\mathrm{n,0}]$ which implies that equation \eqref{eq:TotPbDS} should be integrated in the interval: 
\begin{equation}
\label{eq:TotPbDSProton}
	p(E_\mathrm{n,2}) dE_{\mathrm{n,2}} = dE_{\mathrm{n,2}} \int_{E_\mathrm{n,2}}^{E_\mathrm{n,0}} \dfrac{1}{E_\mathrm{n,0}}\dfrac{1}{E_\mathrm{n,1}} dE_\mathrm{n,1}.
\end{equation}
resulting in:
\begin{equation}
\label{eq:TotPDSsummaryProton}
p(E_\mathrm{n,2}) dE_\mathrm{n,2} = 
\begin{cases}
	\dfrac{1}{E_\mathrm{n,0}} \ln \left( \dfrac{E_\mathrm{n,0}}{E_\mathrm{n,2}} \right)dE_\mathrm{n,2} & \hspace{0.2cm} \mathrm{if~} E_\mathrm{n,2} \in [0, E_\mathrm{n,0}] \\[14pt]
	0  & \hspace{1cm} \mathrm{otherwise.}
\end{cases}
\end{equation}
\begin{figure}
	\begin{center}
  \includegraphics[scale = 0.95]{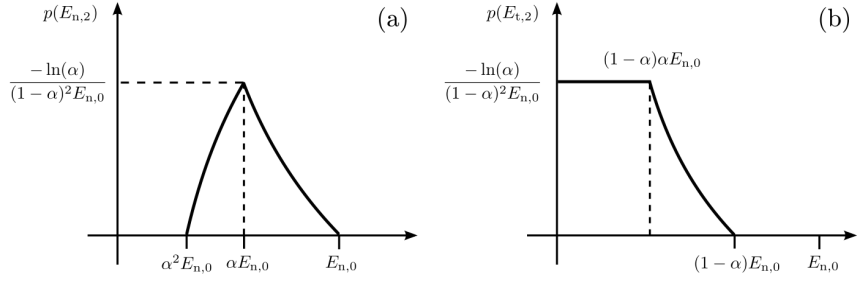}
  	\end{center}
\caption{Probability density functions for the twice scattered neutron (panel (a)) and for the second recoil target (panel (b)) for an incident neutron of energy $E_\mathrm{n,0}$.} 
\label{fig:DoubleScattering}       
\end{figure}

\subsection{Second recoil target energy probability density function}
\label{sec:doubleNP_target}
Consider now the two recoils particles generated from the first and second elastic scattering with a single neutron with energy $E_\mathrm{n,0}$. The probability for the first recoil to have an energy in the range $[E_\mathrm{t,1}, E_\mathrm{t,1} + dE_\mathrm{t,1}]$ is given by equation \eqref{eq:pdfSCt} while the probability of observing the second recoil particle with an energy in the range $[E_\mathrm{t,2}, E_\mathrm{t,2} + dE_\mathrm{t,2}]$ given that after the first collision the neutron has an energy $E_\mathrm{n,1}$ is given by the conditional probability:
\begin{equation}
\label{eq:CondProbTarget}
	p(E_\mathrm{t,2} \mid E_\mathrm{n,1}) dE_\mathrm{t,2} = \dfrac{p(E_\mathrm{t,2} \cap E_\mathrm{n,1})}{p(E_\mathrm{n,1})} dE_\mathrm{t,2} = \dfrac{1}{1 - \alpha} \dfrac{1}{E_\mathrm{n,1}} dE_\mathrm{t,2}.
\end{equation}
The probability of observing the joint events in which the first scattered neutron has energy $E_\mathrm{n,1}$ and collides with a second target transferring to it the energy $E_\mathrm{t,2}$ is:
\begin{equation}
\label{eq:JointProbTarget}
	p(E_\mathrm{t,2} \cap E_\mathrm{n,1}) dE_\mathrm{t,2} = p(E_\mathrm{t,2} \mid E_\mathrm{n,1})p(E_\mathrm{n,1}) dE_\mathrm{t,2}.
\end{equation}
Substitution of the corresponding expressions for the PDFs results in:
\begin{equation}
\label{eq:CondProbTargetExplicit}
	p(E_\mathrm{t,2} \cap E_\mathrm{n,1}) dE_\mathrm{t,2} = \dfrac{1}{(1 - \alpha)^2} \dfrac{1}{E_\mathrm{n,1}}  \dfrac{1}{E_\mathrm{n,0}} dE_\mathrm{n,1} dE_\mathrm{t,2}.
\end{equation}
The probability $p(E_\mathrm{t,2}) dE_\mathrm{t,2}$ of observing a second recoil target with an energy in the range $[E_\mathrm{t,2}, E_\mathrm{t,2} + dE_\mathrm{t,2}]$ is then obtained using the law of total probability, that is, by integrating equation \eqref{eq:CondProbTargetExplicit} over all possible energies $E_\mathrm{n,1}$ that could result in second recoil particle to be in such energy range. 
Note that the second recoil particle can have an energy in the interval $E_\mathrm{t,2} \in [0, (1 - \alpha) \alpha E_\mathrm{n,0}]$ regardless of $E_\mathrm{n,1}$, so the corresponding probability $p(E_\mathrm{t,2}) dE_\mathrm{t,2}$ is:
\begin{align}
\label{eq:TotPb2ndScatt1}
	p(E_\mathrm{t,2}) dE_\mathrm{t,2} & = \int_{\alpha E_\mathrm{n,0}}^{E_\mathrm{n,0}}   \dfrac{1}{(1 - \alpha)^2} \dfrac{1}{E_\mathrm{n,1}}  \dfrac{1}{E_\mathrm{n,0}} dE_\mathrm{n,1} dE_\mathrm{t,2}\\
\label{eq:TotPb2ndScatt1Integrated}	
	& =  \dfrac{1}{(1 - \alpha)^2} \dfrac{1}{E_\mathrm{n,0}} \ln\left( \dfrac{1}{\alpha} \right) dE_\mathrm{t,2}.
\end{align}
\begin{figure}
	\begin{center}
  \includegraphics[scale = 0.6]{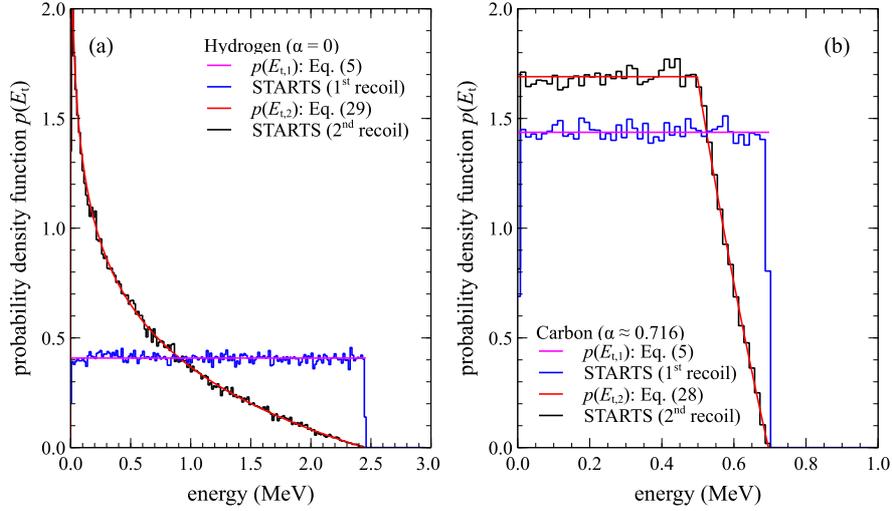}
  	\end{center}
\caption{Probability density functions for the energy of recoil particles for an incident neutron of initial energy of 2.45 MeV undergoing elastic collisions with two particles of the same species: two protons (panel (a)) and two carbon atoms (panel (b)).} 
\label{fig:DoubleScatteringSimulation}      
\end{figure}
If $E_\mathrm{t,2} \in [(1 - \alpha) \alpha E_\mathrm{n,0}, (1 - \alpha) E_\mathrm{n,0}]$ then the incident neutron can only have energies in the interval $[E_\mathrm{t,2}/(1-\alpha), E_\mathrm{n,0}]$ so the corresponding probability $p(E_\mathrm{t,2}) dE_\mathrm{t,2}$ is:
\begin{align}
\label{eq:TotPb2ndScatt2}
	p(E_\mathrm{t,2}) dE_\mathrm{t,2} & = \int_{\tfrac{E_\mathrm{t,2}}{1-\alpha}}^{E_\mathrm{n,0}}   \dfrac{1}{(1 - \alpha)^2} \dfrac{1}{E_{\mathrm{p},1}}  \dfrac{1}{E_{\mathrm{p},0}} dE_{\mathrm{p},1} dE_\mathrm{t,2}\\
\label{eq:TotPb2ndScatt2Integrated}
	 & =  \dfrac{1}{(1 - \alpha)^2} \dfrac{1}{E_\mathrm{n,0}} \ln\left[\dfrac{(1-\alpha) E_\mathrm{n,0}}{ E_\mathrm{t,2}} \right] dE_\mathrm{t,2}.
\end{align}
To summarize, the probability of the second recoil particle to have an energy in the interval $[E_\mathrm{t,2}, E_\mathrm{t,2} + dE_\mathrm{t,2}]$ is given by:
\begin{equation}
\label{eq:TotPDSummary2ndTarget}
	p(E_\mathrm{t,2}) dE_\mathrm{t,2} = 
	\begin{cases}
	\dfrac{1}{(1 - \alpha)^2} \dfrac{1}{E_\mathrm{n,0}} \ln\left( \dfrac{1}{\alpha} \right)  dE_\mathrm{t,2}& \mathrm{if~} E_\mathrm{t,2} \in [0 , (1-\alpha) \alpha E_\mathrm{n,0}]\\[14pt]
	\dfrac{1}{(1 - \alpha)^2} \dfrac{1}{E_\mathrm{n,0}} \ln\left[\dfrac{(1-\alpha) E_\mathrm{n,0} }{ E_\mathrm{t,2}} \right]  dE_\mathrm{t,2} & \mathrm{if~} E_\mathrm{t,2} \in [(1-\alpha) \alpha E_\mathrm{n,0}, (1-\alpha) E_\mathrm{n,0}]\\[14pt]
	0 & \mathrm{otherwise}.
	\end{cases}
\end{equation}
The PDF $p(E_\mathrm{t,2})$ is shown in panel (b) of figure \ref{fig:DoubleScattering}. In the particular case of the two targets being protons, equation \eqref{eq:TotPDSummary2ndTarget} reduces to:
\begin{equation}
\label{eq:TotPDSummary2ndProton}
	p(E_\mathrm{t,2}) dE_\mathrm{t,2} = 
	\begin{cases}
	\dfrac{1}{E_\mathrm{n,0}} \ln\left(\dfrac{E_\mathrm{n,0} }{ E_\mathrm{t,2}  } \right)  dE_\mathrm{t,2} & \mathrm{if~} E_\mathrm{t,2} \in [0, E_\mathrm{n,0}]\\[14pt]
	0 & \mathrm{otherwise}.
	\end{cases}
\end{equation}

Figure \ref{fig:DoubleScatteringSimulation} shows the probability density functions for the energy of two recoil protons and two recoil carbon nuclei calculated according to equations \eqref{eq:pdfSCt} and \eqref{eq:TotPDSummary2ndTarget} with $\alpha = 0$ and $\alpha = 0.716$ (panels (a) and (b) respectively) for $E_\mathrm{n,0} = 2.45$ MeV and compared to STARTS calculations.

\subsection{Total deposited energy for doubly scattered neutrons}
\label{sec:doubleNP_DepEnergy}
Most liquid scintillators are small in size compared to the distance travelled by neutrons with energies in the MeV range in the time required for the ADC to record a few samples (a few nanoseconds). If a neutron was to scatter twice within this time interval then the two recoil particles would be generated on a time scale so short that even present day fast data acquisition system would see the two collisions as a single event with an energy equal to the sum of the energy deposited by each individual recoil particle. As discussed at the beginning of section \ref{sec:doubleNP}, carbon contributes to the light response function only for $L \ll 0.1$ and therefore, even if the energy deposited can be a substantial fraction of the initial neutron energy, it is not discussed further.

\begin{figure}
		\begin{center}
\includegraphics[scale = 0.6]{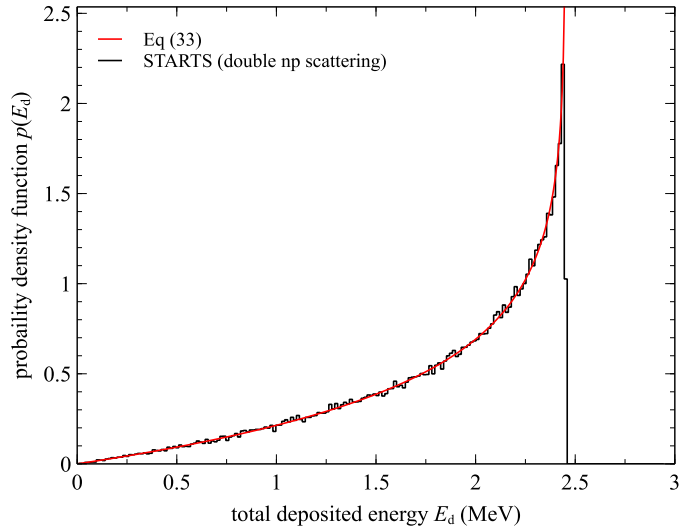}
	\end{center}
\caption{Simulated (red line) and theoretical (dashed black line) probability density function of the total deposited energy by two recoil protons for a neutron with initial energy of 2.45 MeV.} 
\label{fig:DoubleScatteringDepEnergy}      
\end{figure}

Consider instead an incident neutron scattering elastically with two protons: the total energy deposited, in this case, is $E_\mathrm{d} = E_\mathrm{p,1} + E_\mathrm{p,2}$. The probability density function $p(E_\mathrm{d})$ can be obtained by observing that if the first recoil particle deposits an energy $E_\mathrm{p,1}$, then the probability of observing a deposited energy $E_\mathrm{d}$ is equal to the probability of the second recoil proton to have energy $E_\mathrm{p,2}$, that is:
\begin{equation}
\label{eq:DepEnergyCondProb}
	p(E_\mathrm{d} \mid E_\mathrm{p,1}) dE_\mathrm{d} = p(E_\mathrm{p,2} \mid E_\mathrm{p,1}) dE_\mathrm{p,2} = \dfrac{1}{E_\mathrm{n,1}} dE_\mathrm{p,2} = \dfrac{1}{E_\mathrm{n,0} - E_\mathrm{p,1}} dE_\mathrm{p,2}.
\end{equation}
The probability of observing $E_\mathrm{p,1}$ and $E_\mathrm{d}$ is given by:
\begin{equation}
\label{eq:DepEnergyProbGivenEp1}
p(E_\mathrm{p,1} \cap E_\mathrm{d}) dE_\mathrm{d} dE_\mathrm{p,1}  = p(E_\mathrm{p,2} \mid E_\mathrm{p,1})p(E_\mathrm{p,1}) dE_\mathrm{p,2} dE_\mathrm{p,1}.
\end{equation}
Integration over all possible energies $E_\mathrm{p,1}$, recalling that $p(E_\mathrm{p,1}) = 1/E_\mathrm{n,0}$, gives then the probability of observing $E_\mathrm{d}$ \footnotemark:
\begin{align}
\label{eq:DepEnergyProbConv}
	p(E_\mathrm{d}) dE_\mathrm{d} & = \int_0^{E_\mathrm{d}} \dfrac{1}{E_\mathrm{n,0}} \dfrac{1}{E_\mathrm{n,0} - E_\mathrm{p,1}} dE_\mathrm{p,1} dE_\mathrm{p,2} \\
\label{eq:DepEnergyProb}
	& = \dfrac{1}{E_\mathrm{n,0}} \ln \left( \dfrac{E_\mathrm{n,0}}{E_\mathrm{n,0} - E_\mathrm{d}} \right) dE_\mathrm{d} .
\end{align}
\footnotetext{
\label{fn:PDFconv}
Equation \eqref{eq:DepEnergyProbConv} is a special case of the general expression for the PDF of the sum of two continuous random variables $x$ and $y$ with PDFs $p_x$ and $p_y$ which is given by the convolution:
\begin{equation}
	p_{x+y}(z) = p_x \otimes p_y = \int_{-\infty}^{\infty} p_x(\zeta)p_y(z - \zeta) d\zeta. \nonumber
\end{equation}
}
Figure \ref{fig:DoubleScatteringDepEnergy} shows the PDF for the total energy deposited by the two recoil protons from a neutron with $E_\mathrm{n,0} = 2.45$ MeV calculated according to equation \eqref{eq:DepEnergyProb} and with the STARTS code. 
\subsection{Total light output for doubly scattered neutrons}
\label{sec:doubleNP_LightOutput}
\begin{figure}
		\begin{center}
\includegraphics[scale = 0.5]{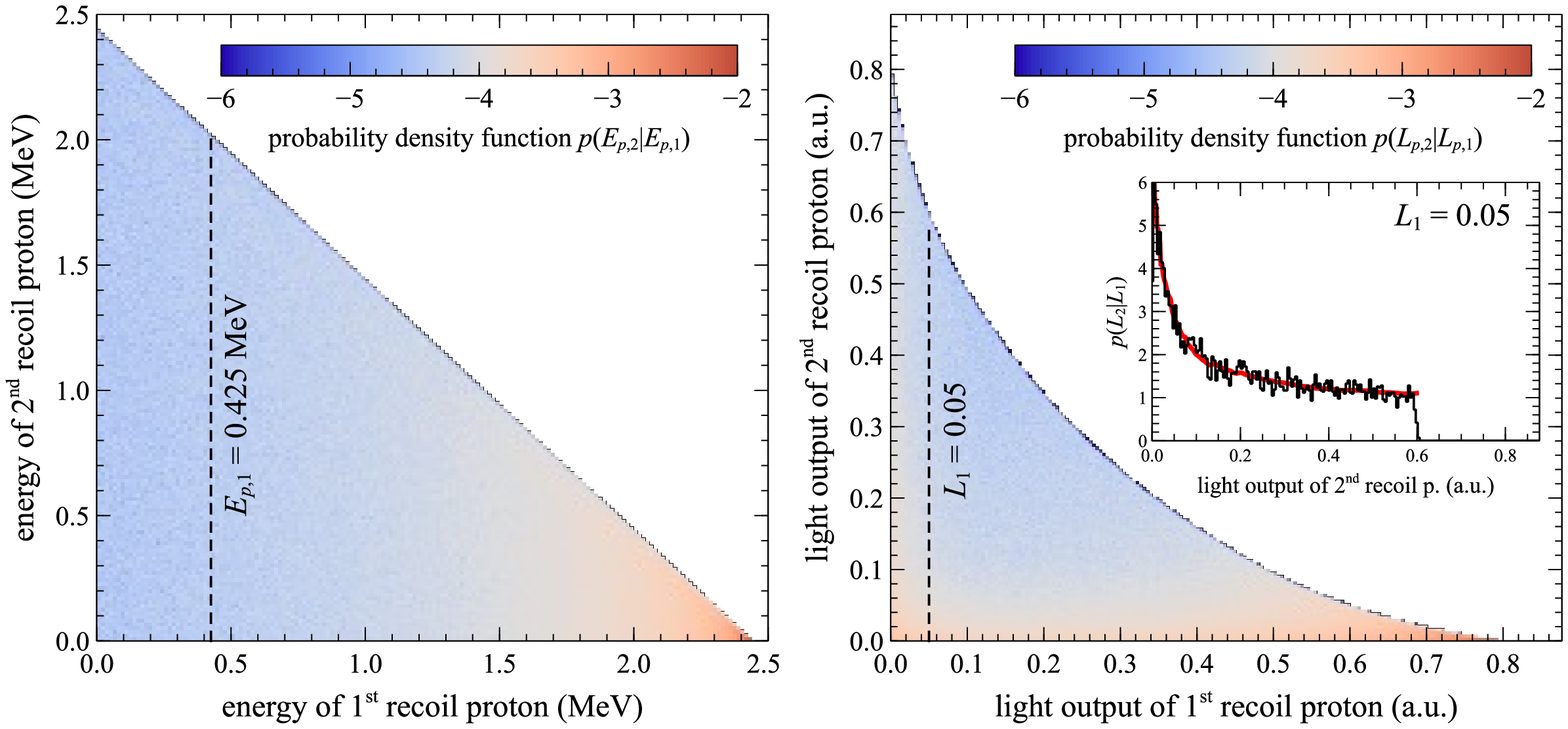}
	\end{center}
\caption{Left panel: PDF $p(E_\mathrm{p,2} \mid E_\mathrm{p,1})$ for the two recoil protons scattered elastically by a neutron with initial energy of 2.45 MeV: the vertical dashed line represents $p(E_\mathrm{p,2} \mid E_\mathrm{p,1} = 0.245 \textrm{~MeV})$. Right panel: PFD  $p(L_2 \mid L_1)$ corresponding to the one based on the energy outcome space shown on the left panel: the vertical dashed line represent the range of possible light outputs $L_2$ given that for $E_\mathrm{p,1} = 0.425$ MeV, $L_1 = 0.05$. The corresponding simulated (black line) and theoretical (red line) PDFs are explicitly shown in the insert.} 
\label{fig:DoubleScatteringSimulation2D}      
\end{figure}

The non-linear relation between the proton recoil energy deposited in the scintillator and the emitted light output transforms non-linearly the two continuous random variables $E_\mathrm{p,1}$ and $E_\mathrm{p,2}$ into the two continuous random variables $L_1$ and $L_2$ thereby transforming non-linearly their outcome space as well. In particular, the outcome space for the total deposited energy by the two recoil protons given by:
\begin{equation}
\label{eq:TotEDep}
	E_\mathrm{p,1} + E_\mathrm{p,2} = E_\mathrm{d}
\end{equation}
becomes 	
\begin{equation}
\label{eq:TotLight}
	L_1 + L_2 = L.
\end{equation}
where if:
\begin{align}
\label{eq:Domain_L1}
	&E_\mathrm{p,1} \in [0, E_\mathrm{n,0}] &  &\Rightarrow & L_1 & \in [0, L(E_\mathrm{n,0})] \\
\label{eq:Domain_L2}	
	&E_\mathrm{p,2} \in [0, E_\mathrm{n,0} - E_\mathrm{p,1}] & &\Rightarrow & L_2 & \in [0, L(E_\mathrm{n,0} - E_\mathrm{p,1})]. 
\end{align}
Figure \ref{fig:DoubleScatteringSimulation2D} shows the effect of this non-linear transformation of the probability density function $p(E_\mathrm{p,2} \mid E_\mathrm{p,1})$ into $p(L_2 \mid L_1)$ for $E_\mathrm{n,0} = 2.45$ MeV calculated by STARTS.
For ``head-on'' collisions, the neutron energy is transferred only to one proton and the light output is $L(E_\mathrm{p}) \approx 0.8$.
For $\theta < \pi$, resulting for example in $E_\mathrm{p,1} = 0.425$ MeV, then $E_\mathrm{p,2} \in [0, 2.025]$ MeV (see left panel of figure \ref{fig:DoubleScatteringSimulation2D}). 
The recoil proton with energy $E_\mathrm{p,1} = 0.425$ MeV will result in a light pulse  $L_1 = 0.05$ and therefore $L_2 \in [0, 0.60]$ (see right panel of figure \ref{fig:DoubleScatteringSimulation2D}). The maximum light output in this case is $L = 0.65$ for a total deposited energy of 2.45 MeV.
The insert on the right panel of figure \ref{fig:DoubleScatteringSimulation2D} shows $p(L_2 \mid L_1 = 0.05)$ calculated by STARTS. 

\begin{figure}
		\begin{center}
\includegraphics[scale = 0.5]{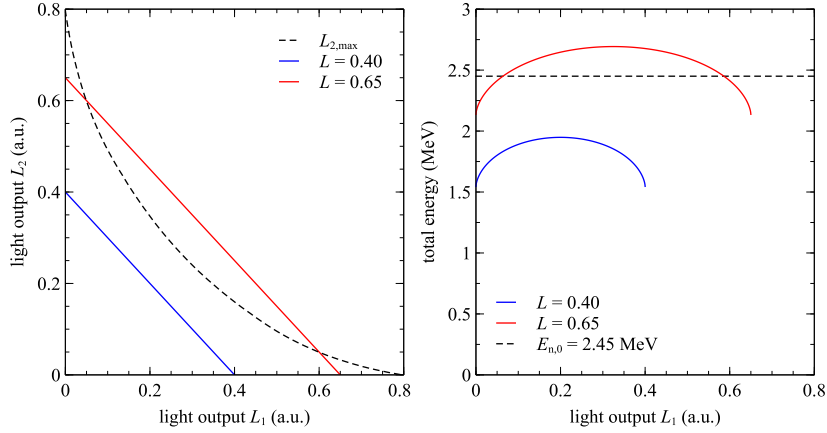}
	\end{center}
\caption{Left panel: maximum possible light output $L_2$ as a function of $L_1$ for $E_\mathrm{n,0} = 2.45$ MeV. The red and blues lines are examples of possible observable light output from two recoil protons. Right panel: total deposited energy as a function of $L_1$ corresponding to the two possible observable light output shown on the left panel.} 
\label{fig:EconsL}      
\end{figure}

Note however that the observable quantity is not the total deposited energy $E_\mathrm{d}$ but the light output $L = L_1 + L_2$ where $L_1$ and $L_2$ must satisfy the condition:
\begin{equation}
\label{eq:Econs}
	E_\mathrm{p}(L_1) + E_\mathrm{p}(L_2) \leq E_\mathrm{n,0}.
\end{equation}
The probability density function for the total light output $p_L$ is then calculated as the convolution $p_{L_1} \otimes p_{L_2}$\footnote{See footnote \ref{fn:PDFconv}.}. The first term, $p_{L_1}$, can be written in terms of $p(E_\mathrm{p,1})$ observing that the probability $p(E_\mathrm{p,1})dE_\mathrm{p}$ of a recoil proton to have an energy in the range $[E_\mathrm{p,1}, E_\mathrm{p,1}+ dE_\mathrm{p,1}]$ is equal to the probability $p(L_1)dL$ that the corresponding light output is in the range $[L_1, L_1+dL_1]$ from which follows\footnotemark:
\begin{equation}
\label{eq:PEeqPL}
p(L_1) = p(E_\mathrm{p,1}) \dfrac{dE_\mathrm{p}(L_1)}{dL_1}.
\end{equation}
\footnotetext{
The quantity $(dE_\mathrm{p}/d\mathrm{L})$ can be calculated numerically from a tabulated data set of $\lbrace E_\mathrm{p}, L \rbrace$ values or analytically if a functional dependence is given as, for example, in  equation \eqref{eq:LO}. In this tutorial, $(dE_\mathrm{p}/d\mathrm{L})$ is calculated numerically using the light output function shown in figure \ref{fig:Verbinski}.} 
Replacing $p(E_\mathrm{p,1})$ with its corresponding expression (see equation \eqref{eq:pdfSCt}), equation \eqref{eq:PEeqPL} becomes:
\begin{equation}
\label{eq:pdfL1}
p(L_1) = \dfrac{1}{E_\mathrm{n,0}} \dfrac{dE_\mathrm{p}(L_1)}{dL_1}.
\end{equation}
In a similar fashion, the PDF for the conditional probability $p(L_2 \mid L_1)dL_2$ is then:
\begin{equation}
\label{eq:pdfL2}
p(L_2 \mid L_1) = \dfrac{1}{E_\mathrm{n,0} - E_\mathrm{p}(L_1)} \dfrac{dE_\mathrm{p}(L_2)}{dL_2}
\end{equation}
and therefore, the probability density function $p(L)$ of observing a total light output $L$ resulting from two recoil protons with light pulses $L_1$ and $L_2$ is: 
\begin{equation}
\label{eq:pdfLconv}
 p(L) = \int_0^{L} \dfrac{1}{E_\mathrm{n,0}} \dfrac{1}{E_\mathrm{n,0} - E_\mathrm{p}(L_1)}  \dfrac{dE_\mathrm{p}(L_1)}{dL_1} \dfrac{dE_\mathrm{p}(L_2)}{dL_2} dL_1.
\end{equation}
The PDFs $p(L_1)$ and $p(L_2 \mid L_1)$ are non-zero only if $L_1$ and $L_2$ satisfy the condition given in equation \eqref{eq:Econs}. This implies that for a given $E_\mathrm{n,0}$ and $L_1$ there is a maximum value $L_{2,\mathrm{max}}$ for which this condition is satisfied and is given by:
\begin{equation}
\label{eq:L2max}
	L_\mathrm{2,max}(L_1) = L[E_\mathrm{n,0} - E_\mathrm{p}(L_1)].	
\end{equation}
The dependence of $L_{2,\mathrm{max}}$ on $L_1$ is shown by the dashed black line in the left panel of figure \ref{fig:EconsL}: the region of possible values of $L_1$ and $L_2$ is the one below the curve $L_{2,\mathrm{max}}(L_1)$. The curve $L_{2,\mathrm{max}}(L_1)$ can be interpreted as all the possible combinations of $L_1$ and $L_2$ values for which $E_\mathrm{d} = E_\mathrm{n,0}$.
As a result, the regions where $p(L_1)$ and $p(L_2 \mid L_1)$ are non-zero depends on both $L$ and $E_\mathrm{n,0}$. Consider in fact first the case where the observed total light output is $L_\mathrm{obs} = 0.4$: such a light output could be obtained by any combination of $L_1$ and $L_2$ values related by equation \eqref{eq:TotLight} (see the blue line in the left panel of figure \ref{fig:EconsL}). The corresponding deposited energy is shown by the blue line in the right panel of figure \ref{fig:EconsL}. Since in this case $E_\mathrm{d} < E_\mathrm{n,0}$, then $p(L_1)$ and $p(L_2)$ are non-zero for all $L_1 \in [0, L_\mathrm{obs}]$.
Conversely, consider now the case in which the observed total light output is $L_\mathrm{obs} = 0.65$ (see red line in the left panel of figure \ref{fig:EconsL}). In this case, $p(L_1)$ and $p(L_2)$ are non-zero only if $L_1 \in [0, L_{1,a}] \cup [L_{1,b}, L_\mathrm{obs}]$ where $L_{1,a}$ and $L_{1,b}$ are the light outputs for which $L_\mathrm{obs} = L_{2,\mathrm{max}}(L_1)$. For $L_1 \in [L_{1,a}, L_{1,b}]$ it turns out that $E_\mathrm{p}(L_1) + E_\mathrm{p}(L_2) > E_\mathrm{n,0}$ (as shown by the red line on the right panel of figure \ref{fig:EconsL}) which is of course not physically possible.
It is clear then, that as $L$ increases so does the integral $p(L)$ until $L = L_\mathrm{B,1}$ with:
\begin{equation}
\label{eq:LB1}
	L_\mathrm{B,1} = L_1^* + L_\mathrm{2, max}(L_1^*)
\end{equation} 
where $L_1^*$ is the point of tangency, i.e.:
\begin{equation}
\label{eq:tangency}
	\dfrac{d L_\mathrm{2, max}(L_1)}{d L_1} \bigg\vert_{L_1 = L_1^*} = -1.
\end{equation}
\begin{figure}
		\begin{center}
	\includegraphics[scale = 0.6]{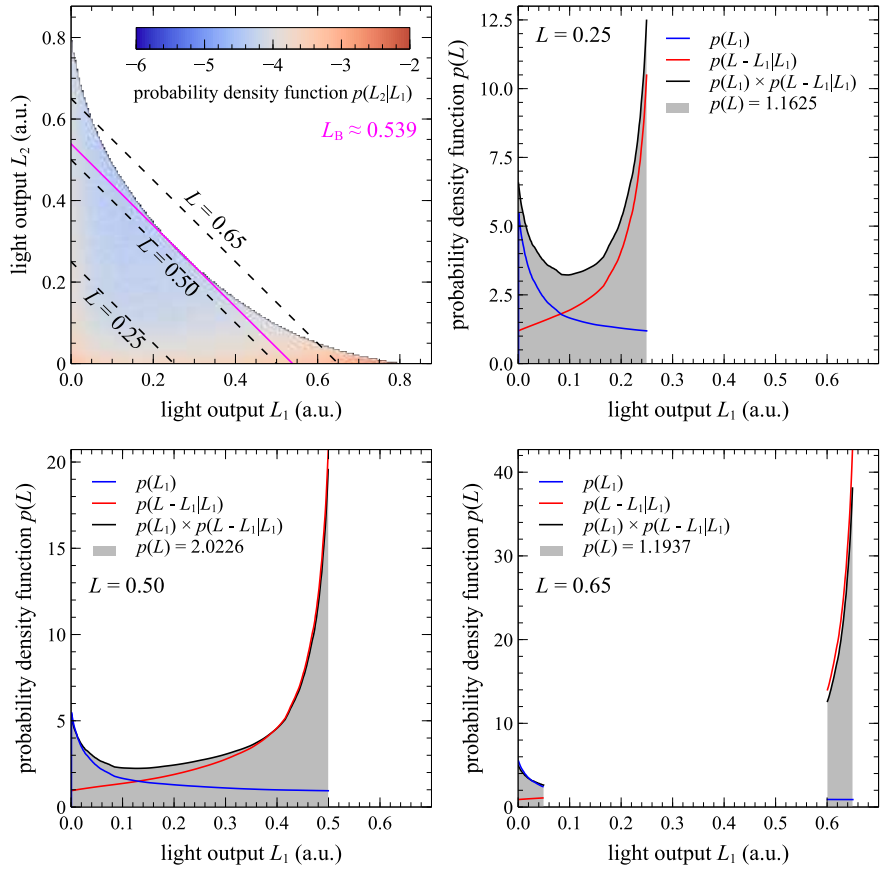}
		\end{center}
	\caption{Evaluation of the PDF $p(L)$ for the total light output $L$ for $L = 0.25$, $L = 0.50$ and $L = 0.65$.} 
	\label{fig:Convolution}      
\end{figure}
The PDF $p(L)$ reaches its maximum for $L = L_\mathrm{B,1}$ and goes to zero as $L \rightarrow 0$ since the regions where the PDF $p(L_1)$ and $p(L_2)$ are non-zero become vanishing small\footnote{In the case of the light output function used in this tutorial and for $E_\mathrm{n,0} = 2.45$ MeV, $L_1^* \approx 0.2697$ and $L_\mathrm{B,1} \approx 0.5389$.}. 
The way in which the convolution integral in equation \eqref{eq:pdfLconv} is calculated as a function of $L$ is elucidated in figure \ref{fig:Convolution} which shows how the integrand depends on $L_1$ for three different values of the total light output $L$.
Figure \ref{fig:nppLO} shows instead the PDF $p(L)$ given by equation \eqref{eq:pdfLconv} for all possible values of the total light output, that is for $L \in [0, L_\mathrm{M}]$,  compared with the results from STARTS.
\begin{figure}
		\begin{center}
  \includegraphics[scale = 0.6]{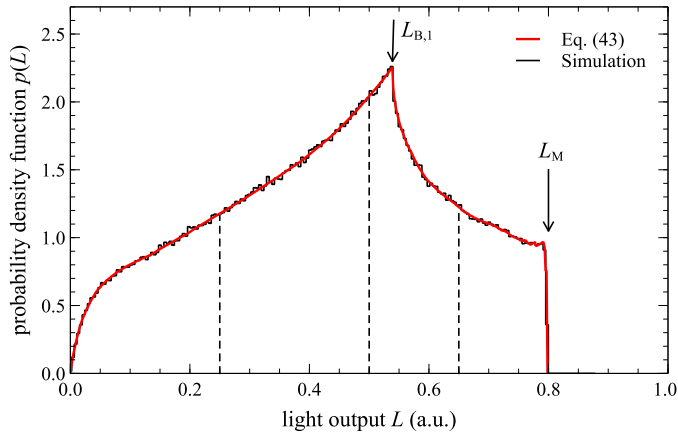}
  	\end{center}
\caption{Probability density function for the light output corresponding to the total energy deposited in the scintillator when a neutron with an energy of 2.45 MeV scatters elastically with two protons: simplified simulation (black) and expected (red). The vertical dashed lines indicate the values of the total light output shown in figure \ref{fig:Convolution}.} 
\label{fig:nppLO}      
\end{figure}

\begin{figure}
		\begin{center}
  \includegraphics[scale = 0.6]{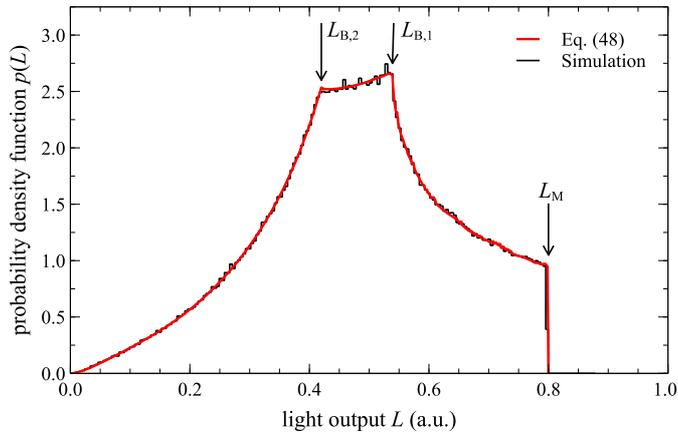}
  	\end{center}
\caption{Probability density function for the light output corresponding to the total energy deposited in the scintillator when a neutron with an energy of 2.45 MeV scatters elastically with three protons: simplified simulation (black) and expected (red).} 
\label{fig:npppLO}      
\end{figure}

The predicted contribution to the total light output response function due to a neutron of a given initial energy making two elastic scattering with protons shown in figure \ref{fig:nppLO} can be individuated in the NRESP response function shown in figure \ref{fig:LRFEp}: the sharp knee for $L \approx 0.539$ can now be understood in terms of the maximum energy repartition between the two recoil protons. 
A closer comparison between figures \ref{fig:LRFEp} and \ref{fig:nppLO} reveals however that, for $L < L_\mathrm{B,1}$, $p(L)$ does not drop as predicted by equation \eqref{eq:pdfLconv} which indicates that double scattering on protons is not sufficient to reproduce NRESP response function. For this to happen, it is necessary to consider the contribution from triple np scattering.  
In a fashion similar to what has been done for the double scattering case, it is possible to write the PDF for the light output for the third recoil proton as:
\begin{equation}
\label{eq:TriplePDF}
	p(L_3) = \dfrac{1}{E_\mathrm{n,0} - E_\mathrm{p}(L_1) - E_\mathrm{p}(L_2)} \dfrac{dE_\mathrm{p}(L_3)}{dL_3}.
\end{equation}
The probability density function of the sums $L = L_1 + L_2 + L_3$ is then calculated as the convolution $p_L = p_{L_1} \otimes p_{L_2} \otimes p_{L_3}$: 
\begin{equation}
\label{eq:TripleCNV}
	p(L) = \int_0^L p(L_1) \int_0^{L-L_1} p(L_2) p(L - L_1 - L_2) dL_2 dL1.
\end{equation}
and is shown in figure \ref{fig:npppLO} together with the one calculated by STARTS. As can be seen, $p(L)$ does not drop much for $L \in [L_\mathrm{B,2}, L_\mathrm{B,1}]$ where $L_\mathrm{B,2}$ corresponds to situation in which $E_\mathrm{p}(L_1) + E_\mathrm{p}(L_2) + E_\mathrm{p}(L_3) = E_\mathrm{n,0}$.
It is clear from the results shown in figures \ref{fig:LRFEp}, \ref{fig:nppLO} and \ref{fig:npppLO} that a linear combination of the response functions from one, two and three recoil protons can reproduce NRESP output. 
This is however postponed until section \ref{sec:Comparison} as the last important component to the total light response function, that is the one arising from recoil protons from a neutron that has undergone a prior elastic collision with a carbon atom (track ``d'' of figure \ref{fig:DoubleScattering}), is discussed in the next section. 

\section{Response function for neutron scattering with different targets}
\label{sec:mixedNP}

From the discussion in sections \ref{sec:singleNP} and \ref{sec:doubleNP_target} it is clear that, even if the light output from the recoil carbon makes a negligible contribution to the response function, the energy repartition between the recoil and scattered particles will affect the light output of the scattered proton in the second elastic collision. 
The corresponding energy and light output PDFs can be derived by generalizing to the case where an incident neutron with initial energy $E_\mathrm{n,0}$ makes two elastic collisions with atoms characterized by $A_1$ and $A_2$ such that $A_1 > A_2$ (and therefore $\alpha_1 > \alpha_2$). 
The probability of observing a neutron with energy between $[E_\mathrm{n,1}, E_\mathrm{n,1} + dE_\mathrm{n,1}]$ after the scattering with $A_1$ is:
\begin{equation}
\label{eq:mix_EnA1}
	p(E_{\mathrm{n},1}) dE_{\mathrm{n},1} = \dfrac{1}{1-\alpha_1}\dfrac{1}{E_{\mathrm{n},0}} dE_{\mathrm{n},1}.
\end{equation}
The probability of the twice scattered neutron to have an energy in the range $[E_\mathrm{n,2}, E_\mathrm{n,2} + dE_\mathrm{n,2}]$ after the scattering with $A_2$ is obtained by applying the law of total probability to the joint event $p(E_\mathrm{n,2} \cap E_\mathrm{n,1}) = p(E_\mathrm{n,2} \mid E_\mathrm{n,1}) p(E_\mathrm{n,1})$ where the integral is carried out over all possible $E_\mathrm{n,1}$:
\begin{equation}
\label{eq:mix_EnA2}
	p(E_{\mathrm{n},2}) dE_{\mathrm{n},2} = \dfrac{1}{1-\alpha_1} \dfrac{1}{1-\alpha_2}  \dfrac{dE_{\mathrm{n},2}}{E_{\mathrm{n},0}}  \int \dfrac{1}{E_{\mathrm{n},1}} dE_{\mathrm{n},1}
\end{equation}
The integration limits above depends on $E_\mathrm{n,2}$ possible ranges. In particular
\begin{align}
\label{eq:mix_En2GivenEn1}
	\mathrm{if~} 
	\begin{dcases}
		E_{\mathrm{n}, 2} \in [\alpha_1 E_{\mathrm{n}, 0}, E_{\mathrm{n}, 0}] & \Rightarrow  E_{\mathrm{n}, 1} \in [E_{\mathrm{n}, 2}, E_{\mathrm{n}, 0}] \\[6pt]
		E_{\mathrm{n}, 2} \in [\alpha_2 E_{\mathrm{n}, 0}, \alpha_1 E_{\mathrm{n}, 0}] & \Rightarrow E_{\mathrm{n}, 1} \in [\alpha_1 E_{\mathrm{n}, 0}, E_{\mathrm{n}, 0}] \\[6pt]
		E_{\mathrm{n}, 2} \in [\alpha_1 \alpha_2 E_{\mathrm{n}, 0}, \alpha_2 E_{\mathrm{n}, 0}] & \Rightarrow E_{\mathrm{n}, 1} \in [\alpha_1 E_{\mathrm{n}, 0}, E_{\mathrm{n}, 2}/\alpha_2]. 				
	\end{dcases}
\end{align}
then:
\begin{align}
\label{eq:mix_Et2PDFE1}
	p(E_{\mathrm{n},2}) dE_{\mathrm{n},2} & = \dfrac{1}{1-\alpha_1} \dfrac{1}{1-\alpha_2}  \dfrac{dE_{\mathrm{n},2}}{E_{\mathrm{n},0}} \times\\
	&
	\begin{dcases}
		\ln \left(\dfrac{E_{\mathrm{n},0}}{E_{\mathrm{n},2}} \right) &\mathrm{if~} E_{\mathrm{n}, 2} \in [\alpha_1 E_{\mathrm{n}, 0}, E_{\mathrm{n}, 0}] \\[6pt]
		\ln \left(\dfrac{1}{\alpha_1} \right) &\mathrm{if~}  E_{\mathrm{n}, 2} \in [\alpha_2 E_{\mathrm{n}, 0}, \alpha_1 E_{\mathrm{n}, 0}]\\[6pt]
		\ln \left(\dfrac{E_{\mathrm{n},2}}{\alpha_1 \alpha_2 E_{\mathrm{n},0}} \right) &\mathrm{if~} E_{\mathrm{n}, 2} \in [\alpha_1 \alpha_2 E_{\mathrm{n}, 0}, \alpha_2 E_{\mathrm{n}, 0}]\\[6pt]
		0 &\mathrm{otherwise}.
	\end{dcases}\nonumber
\end{align}
In the case $\alpha_1 > 0$ and $\alpha_2 = 0$, the PDF $p(E_\mathrm{n,2})$ then becomes:
\begin{align}
\label{eq:mix_Et2PDFE1}
	p(E_{\mathrm{n},2}) dE_{\mathrm{n},2} = \dfrac{1}{1-\alpha_1} \dfrac{dE_{\mathrm{n},2}}{E_{\mathrm{n},0}} \times
	\begin{dcases}
		\ln \left(\dfrac{E_{\mathrm{n},0}}{E_{\mathrm{n},2}} \right) &\mathrm{if~} E_{\mathrm{n}, 2} \in [\alpha_1 E_{\mathrm{n}, 0}, E_{\mathrm{n}, 0}] \\[6pt]
		\ln \left(\dfrac{1}{\alpha_1} \right) &\mathrm{if~}  E_{\mathrm{n}, 2} \in [0, \alpha_1 E_{\mathrm{n}, 0}] \\[6pt]
		0 &\mathrm{otherwise}.
	\end{dcases}
\end{align}

\begin{figure}
	\begin{center}
  \includegraphics[scale = 0.6]{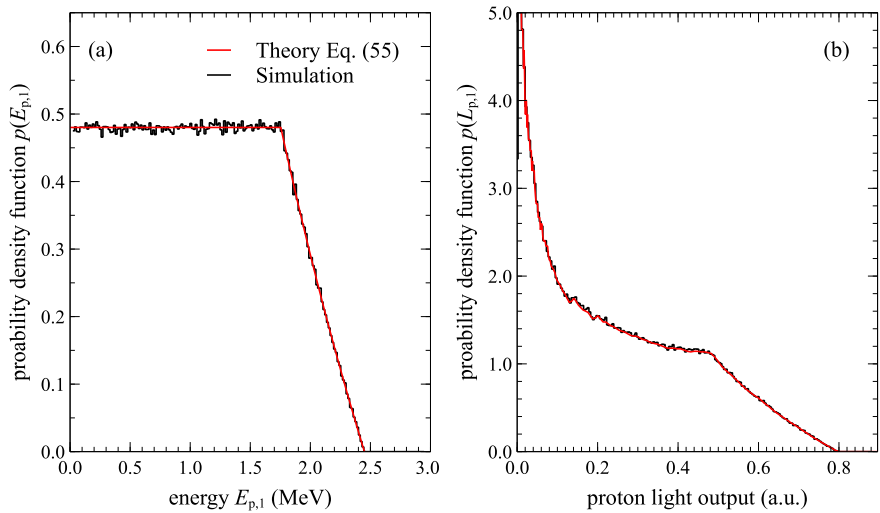}
  	\end{center}
\caption{Simulated (black line) and theoretical (red line) probability density function for the energy (panel (a)) and the light output (panel (b)) for the recoil proton from an elastic scattering with a neutron that has previously undergone a collision with a carbon atom.} 
\label{fig:nCp}      
\end{figure}

The probability density function for the energy of the recoil target $A_1$ is given by \eqref{eq:pdfSCt} with $\alpha$ replaced by $\alpha_1$ if $E_{A_1,1} \in [0, (1-\alpha_1) E_{\mathrm{n},0}]$ and zero everywhere else. 
The PDF for the energy of recoil target $A_2$ is given by equation \eqref{eq:mix_EnA2} but the integration limits are now given by:
\begin{align}
\label{eq:mix_Et2PDFconditions}
	\mathrm{if~}
	\begin{dcases}
		E_{{A_2}                                                                                                                  , 1} \in [0, (1-\alpha_2) \alpha_1 E_{\mathrm{n}, 0}] &\Rightarrow E_{\mathrm{n}, 1} \in [\alpha_1 E_{\mathrm{n}, 0}, E_{\mathrm{n}, 0}] \\[6pt]
		E_{{A_2}, 1} \in [(1-\alpha_2)\alpha_1 E_{\mathrm{n}, 0}, (1-\alpha_2) E_{\mathrm{n}, 0}] &\Rightarrow E_{\mathrm{n}, 1} \in \left[ \dfrac{E_{A_2,1}}{1-\alpha_2}, E_{\mathrm{n}, 0} \right]. 			
	\end{dcases}	
\end{align}
In this case, $(1-\alpha_2) E_{\mathrm{n}, 0}$ corresponds to $A_2$'s maximum possible energy if the neutron has lost no energy in the collisions with $A_1$ (grazing collision) while $ (1-\alpha_2) \alpha_1 E_{\mathrm{n}, 0}$ is $A_2$'s maximum possible energy if the
neutron has lost the maximum energy possible in a ``head-on'' collision with $A_1$. 
Integration of equation \eqref{eq:mix_EnA2} with $E_\mathrm{n,1}$ in the intervals specified in \eqref{eq:mix_Et2PDFconditions} gives the probability for the second recoil target to have an energy in $[E_{A_2,1}, E_{A_2,1} + dE_{A_2,1}]$:
\begin{align}
\label{eq:mix_Et2PDF}
	p(E_{A_2,1}) dE_{A_2,1}  &= \dfrac{1}{1-\alpha_1} \dfrac{1}{1-\alpha_2}  \dfrac{dE_{A_2,1}}{E_{\mathrm{n},0}} \times\\
	&
	\begin{dcases}
		\ln \left(\dfrac{1}{\alpha_1} \right) &\mathrm{if} \; E_{{A_2}                                                                                                                  , 1} \in [0, (1-\alpha_2) \alpha_1 E_{\mathrm{n}, 0}] \\[6pt]
		\ln \left[\dfrac{E_{\mathrm{n},0}}{E_{A_2,1}} (1-\alpha_2) \right] &\mathrm{if} \; E_{{A_2}, 1} \in [(1-\alpha_2)\alpha_1 E_{\mathrm{n}, 0}, (1-\alpha_2) E_{\mathrm{n}, 0}]\\[6pt]
		0 &\mathrm{otherwise}.
	\end{dcases}\nonumber
\end{align}
In the case $\alpha_1 > 0$ and $\alpha_2 = 0$, the probability of observing the recoil proton with energy in the range $[E_\mathrm{p,1},E_\mathrm{p,1} + dE_\mathrm{p,1}]$ is therefore:
\begin{align}
\label{eq:mix_pPDF}
	p(E_\mathrm{p,1}) dE_\mathrm{p,1}  = \dfrac{1}{1-\alpha_1} \dfrac{dE_\mathrm{p,1}}{E_{\mathrm{n},0}} \times
	\begin{dcases}
		\ln \left(\dfrac{1}{\alpha_1} \right) &\mathrm{if} \; E_\mathrm{p,1} \in [0, \alpha_1 E_\mathrm{n,0}]\\[6pt]
		\ln \left(\dfrac{E_{\mathrm{n},0}}{E_\mathrm{p,1}} \right) &\mathrm{if} \; E_\mathrm{p,1} \in [\alpha_1 E_\mathrm{n,0}, E_\mathrm{n,0}]\\[6pt]
		0 &\mathrm{otherwise}.
	\end{dcases} 
\end{align}
Panel (a) of figure \ref{fig:nCp} shows the PDF for the energy of the recoil proton as a function of $E_\mathrm{p,1}$ calculated by equation \eqref{eq:mix_pPDF} together with STARTS results. The corresponding PDF for the light output is given by applying equation \eqref{eq:NLOSnp} to equation \eqref{eq:mix_pPDF} and is shown in panel (b).

\section{Comparison with full Monte Carlo simulations}
\label{sec:Comparison}
It is now possible to proceed to the qualitative comparison between a combination of the theoretical PDFs derived in the previous sections with the light output response function obtained by full Monte Carlo simulation such as NRESP. 
In this context, a full Monte Carlo simulation refers to a simulation in which the proper cross-sections of the relevant processes are taken into account, the the density of hydrogen and carbon atoms in the scintillation material are specified and a realistic geometrical model of the detector is used. 
Without this level of details, neither STARTS nor the analytical approach can properly estimate the relative importance (amplitude) of the different components constituting the light output response function although one can reasonably assume that the role of multiple np scattering becomes more important as the size of the detector increases and that collisions with carbon nuclei followed by collisions on hydrogen are always present given that the cross-sections for these processes are of the same order of magnitude. 
The weights of the light output response functions corresponding to the different elastic scattering processes are obtained by a simple multiple linear regression where the dependent variable is the overall response function calculated by NRESP.  
The relative amplitude of the different components so obtained has just an indicative (qualitative) nature and for a proper estimate of these weights one should exclusively use full Monte Carlo codes.
\begin{figure}
		\begin{center}
  \includegraphics[scale = 0.45]{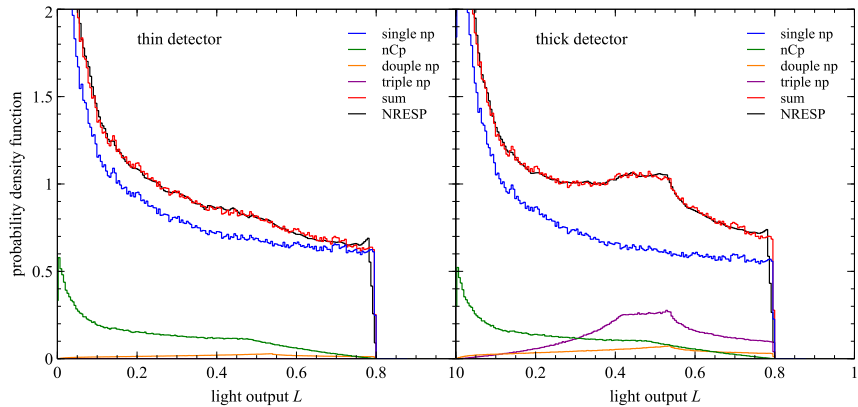}
  	\end{center}
\caption{Comparison between the light output response function of a ``thin'' (left panel) and a ``thick'' (right panel) liquid scintillator to a mono-energetic neutron with energy of 2.45 MeV calculated by NRESP (black line) and the predicted one (red line) from a linear regression of the theoretical light output response function for single (blue line), double (orange line) and triple (dark magenta) np scattering events together with the nCp scattering contribution (green line).}
\label{fig:LORFmatch}       
\end{figure}

With these words of caution well present in mind, the comparison between the NRESP light output response function and the one from the linear regression based on STARTS calculations is shown in figure \ref{fig:LORFmatch} for both the ``thin'' and ``thick'' detectors.
As can be seen, the light response function for the ``thin'' detector can be well matched neglecting the contribution from triple np scattering while this component is essential to match the response function for the ``thick detector''. 

\begin{figure}
	\begin{center}
  \includegraphics[scale = 0.6]{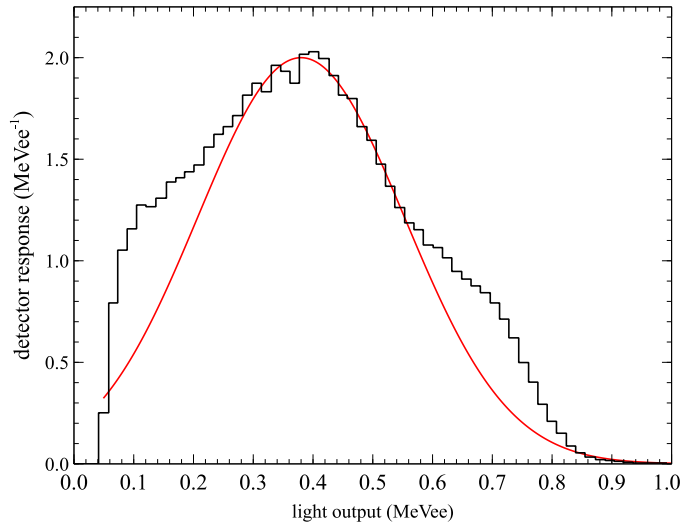}
  	\end{center}
\caption{An example of the central limit theorem: multiple np scattering giving rise to a normal-like distribution contribution (red line) to the response function for 2.45 MeV neutrons into a large EJ-309 liquid scintillator (black line): data from \cite{Enqvist}. The experimental data have been acquired with an acquisition threshold which explain the reason why the response function goes to zero at low light output.} 
\label{fig:Enqvist}      
\end{figure}

As an example of how the analytical approach can be used to make predictions regarding the expected response function of a liquid scintillator, consider the case of a scintillator with a very large active volume. It is obvious that multiple elastic scattering of the incident neutron on protons should make a large contribution to the total light output response functions. Observing that the response function for multiple np scattering can be calculated by multiple convolutions of the same probability density function, although each weighted by a different factor which depends on the scattered neutron energy before each collision, and using the central limit theorem\footnotemark{} then it is possible to predict that expected response function should approximate a normal probability density function. This is indeed the case as shown in figure \ref{fig:Enqvist} where the detector response function of a 12.7-by-12.7 cm cylindrical EJ-309 liquid scintillator measured experimentally \cite{Enqvist} is compared with a normal distribution. 

\footnotetext{The central limit theorem holds only if both the average value and variance of the random variable exist. The random variable in this case is the total light output $L$ for which the existence of its average value and variance is guaranteed.}

Having now described, albeit qualitatively, the building blocks of the light response function of a liquid scintillator in the simplest case, it is now possible to tackle its interpretation in presence of those effects neglected so far and in more realistic scenarios such as those encountered in fusion devices. 

\section{Response functions in realistic scenarios}
\label{sec:complications}
In this section additional aspects affecting the response function of liquid scintillators that have been neglected in the previous sections are discussed. These can be divided in two categoties: those aspects that are intrinsic to the scintillator properties which are discussed in sections \ref{sec:DetectorOthers} and \ref{sec:efficiency} and those which depend on the incident neutron energy spectrum, discussed in section \ref{sec:FusionSpectra}. 

\subsection{Detector properties affecting the light output response function}
\label{sec:DetectorOthers}
The response function discussed so far has been limited to the case of 2.45 MeV mono-energetic incident neutrons for which the predominant interaction mechanism is elastic scattering on hydrogen and carbon nuclei. 
For neutrons of higher energies the following competing processes occur: inelastic nC scattering (for $E_\mathrm{n} > 4.8$  MeV) and the nuclear reactions $^{12}\mathrm{C}(\mathrm{n}, \alpha){^9}\mathrm{Be}$ (for $E_\mathrm{n} > 7.4$  MeV) and $^{12}\mathrm{C}(\mathrm{n}, \mathrm{n}'){^9}\mathrm{Be^*}$ (for $E_\mathrm{n} > 10$ MeV). 
At these high energies, nuclear reactions induced by the neutron interacting with the detector material surrounding the liquid scintillator cell are also possible. 
Inelastic scattering results in the scattered neutron and recoil particles to have a lower energy than in the case of elastic scattering. The light output from these recoils particles have smaller amplitude and will be distributed continuously from zero to the maximum light output possible.
The light output from $\alpha$-s of few MeV is much higher than that of recoil carbons of similar energy but still lower by a factor of about 10 compared to the one for recoils protons (see figure \ref{fig:Verbinski}). The contribution from $\alpha$-s appear as an additional ``edge'' at the low end part of the light output response function superimposed to the ``box-like'' response for protons. Indicating with $L_\mathrm{p}$ the light output resulting from a ``head-on'' collision of a high energy neutron with a proton and with $L_\alpha$ the light output produced by an $\alpha$ particle then $L_\alpha/L_\mathrm{p} \approx 0.1$ for $E_\mathrm{n,0} = 15$ MeV \cite{Klein}. Due to the complexity of these competing events as well as the strong angular dependence of their cross section, full Monte Carlo simulations are indispensable for a correct modelling of experimental response functions at neutron energies above few MeVs.

The transport of the scintillation light from its point of generation inside the active detector volume to the PMT becomes important for large detectors or detectors with one dimension much larger than the other. 
This is due to the fact that the scintillator itself is a self-absorbing medium and therefore the same deposited energy will result in two light pulses of different amplitude depending on the location of the recoil particle generation with respect to the PMT. 
The effect of light transport and self-attenuation is usually negligible for small volume liquid scintillators (a few centimetres). 
For larger scintillators, self-attenuation results in light pulses being distributed at lower light output than expected. For example, in the case of incident mono-energetic neutrons onto a long cylindrical scintillator with a PMT at one end of the cylinder, self-absorption will result in the ``smearing'' of the ``head-on'' collision edge towards lower amplitudes thus worsening the detector energy resolution. If liquid scintillators are to be used for neutron spectroscopy it is therefore essential that they are of limited dimensions. 
Codes such GEANT4 \cite{Agostinelli} and  MCNP-PoliMI \cite{Pozzi} are capable of treating properly the transport of photons and should be used for scintillators of shape more complex than the standard cylindrical one and for large volumes. 

Two additional quantities that affect the light response function are the light output as a function of the deposited energy by different recoil particles and the energy resolution function. 
The light output function depends by the size and geometry of the scintillator and affects the relative position of the different features in the light output pulse height spectrum and in particular the position of the ``edge'' corresponding to the deposition of the full neutron energy in a single elastic scattering with a proton (``head-on'' collision). 
The energy resolution dependency on the neutron energy results in the smearing of the response function and it is mostly predominant at the location of the ``head-on edge''. 
The energy resolution is affected in turn by the geometry and size of the scintillator and typically ranges between 5 and 30 \% in the neutron energy range 0 - 3 MeV.
For a discussion on the different methods for measuring the light output and energy resolution functions the reader is referred to \cite{Enqvist, Klein} the references therein. Since the ``head-on edge'' region is the one generated by neutrons that have not undergone any collision, the spectral information on the neutron source is mainly contained in this region and therefore the determination of the light output and energy resolution functions is very important especially at the energy of interest (for example at 2.45 and 14.1 MeV for fusion plasmas).

\subsection{Detector efficiency}
\label{sec:efficiency}
The efficiency of a liquid scintillator to a mono-energetic neutron with energy $E_\mathrm{n,0}$ for single elastic scattering on hydrogen is given by \cite{Knoll}:
\begin{equation}
\label{eq:eff_single}
	\epsilon(E_\mathrm{n,0}) = \dfrac{n_\mathrm{H} \sigma_\mathrm{H}(E_\mathrm{n,0})}{n_\mathrm{H} \sigma_\mathrm{H}(E_\mathrm{n,0}) + n_\mathrm{C} \sigma_\mathrm{C}(E_\mathrm{n,0})} \left\lbrace 1- e^{ - \left[ n_\mathrm{H} \sigma_\mathrm{H}(E_\mathrm{n,0}) + n_\mathrm{C} \sigma_\mathrm{C}(E_\mathrm{n,0}) \right] d } \right\rbrace
\end{equation}
where $d$ is the detector thickness, $n_\mathrm{H}$ and $n_\mathrm{C}$ are the number density of hydrogen and carbon atoms respectively, and $\sigma_\mathrm{H}$ and $\sigma_\mathrm{C}$ are the neutron scattering cross-section on hydrogen and carbon. The exponential terms represents the fraction of neutrons of energy $E_\mathrm{n,0}$ passing through a scintillator of thickness $d$ (i.e. the uncollided neutrons) and therfore the terms within curly brackets represents the fraction of neutrons interacting in the scintillation material. 

\begin{figure}
	\begin{center}
  \includegraphics[scale = 0.6]{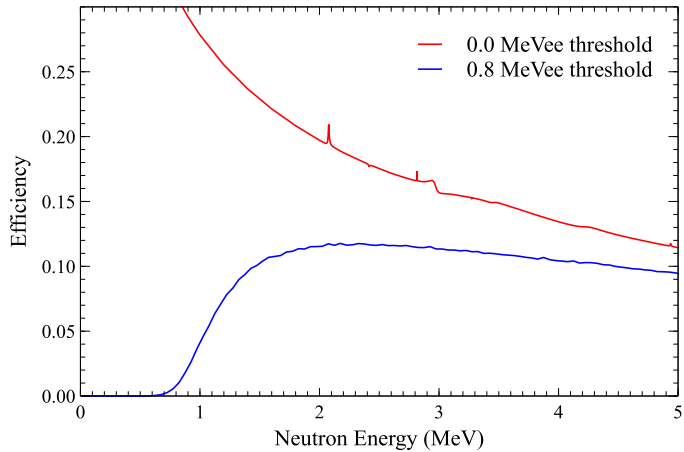}
  	\end{center}
\caption{Liquid scintillator efficiency calculated by NRESP for a 1.5 cm thick scintillator with and without an acquisition threshold.} 
\label{fig:efficiency}      
\end{figure}

The detector efficiency can be calculated using full Monte Carlo codes as the integral of the pulse height spectrum: the efficiency of a 1.5 cm thick scintillator as a function of the incident neutron energy is shown in figure \ref{fig:efficiency}. Since in reality, pulse height spectra are measured with an acquisition threshold, as shown in figure \ref{fig:Enqvist}, the lower limit of the integration must be accurately determined.  
This is typically done by calibrating the detector light output against the energy deposited by recoil electron using standard $\gamma$-rays calibration sources (such as $^{22}$Na, $^{137}$Cs and $^{207}$Bi) and then converting the experimental light output threshold into a deposited energy threshold using the corresponding light output function \cite{Ecal, Fowler}. The light output so calibrated is given in units of MeV electron-equivalent (MeVee) as shown in figure \ref{fig:Enqvist}. In this case, the light output threshold is approximately 0.1 MeVee resulting in a deposited energy threshold of approximately 0.8 MeVee for recoil protons (see figure \ref{fig:Verbinski}). In particular, the efficiency and the scintillator cross-sectional area determine the number of counts per seconds that are measured for a given incident neutron flux.
As shown in figure \ref{fig:efficiency}, the probability of a neutron to interact with the scintillator increases as its energy decreases. As a results, neutrons that have undergone (multiple) scattering before reaching the detector can make a significant contribution to the pulse height spectrum. Neutrons that have collided first with the scintillation material will have an increased probability of making multiple collisions especially for thick detectors as the detection efficiency increases.
It is clear then that the dependency of the efficiency on the neutron energy has a non-trivial effect on the light response function both in shape and amplitude. Full Monte Carlo codes ensure that all these effects are included in the response function.

\subsection{Neutron energy spectra in fusion devices and forward modelling}
\label{sec:FusionSpectra}
In fusion devices, liquid scintillators are typically located inside thick neutron shield and view the neutron source (the plasma) via long collimators and the energy spectrum of the neutrons at the detector are anything but mono-energetic.
A typical fusion neutron energy spectrum contains spectral components from \emph{i}) DD and DT thermal emission (approximately gaussian with a width which depends on the plasma ion temperature), \emph{ii}) fusion reactions involving ions with energies much larger than the thermal ions produced by additional heating and \emph{iii}) scattered neutrons, that is, neutrons that before their first interaction with the scintillator have already suffered one or multiple collisions with the material surrounding it (from the fusion device itself, to the neutron diagnostic in which the detector is installed). 
\begin{figure}
	\begin{center}
  \includegraphics[scale = 0.6]{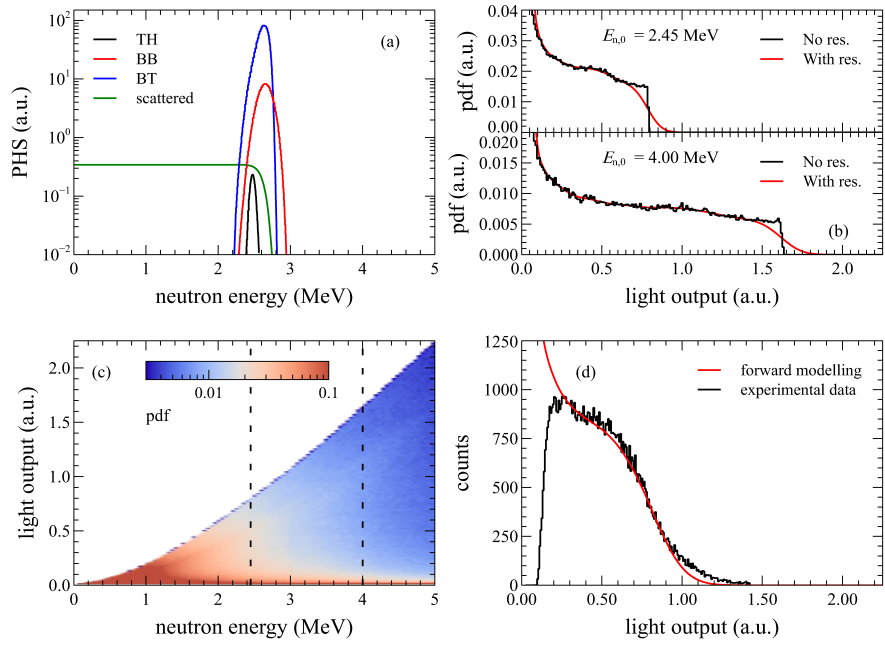}
  	\end{center}
\caption{Example of forward modelling. Panel (a) shows the individual components of the neutron energy spectrum for a neutral beam heated DD fusion plasmas at the detector in its different components: THermal (TH), Beam-Beam (BB) and Beam-Thermal (BT) assuming a 5 \% contribution from scattered neutrons. Panel (b) shows light output response function matrix: the light output response function for two specific energies (2.45 and 4.00 MeV) indicated by the vertical dashed black lines are shown in panels (c) and (d) with and without the effect of the finite energy resolution. Panel (e) shows the folding of the energy spectrum in panel (a) with the response function matrix in panel (b). Finally, in panel (f) the linear combination of the folded response functions for all neutron energies shown in panel (e) are compared with the experimentally measured light output pulse height spectrum.} 
\label{fig:FM}      
\end{figure}
In such a situation, the interpretation of the experimental response function requires the accurate estimate of the neutron flux and energy spectrum at the detector location. In turns this requires accurate plasma physics models for the calculation of the DD and DT neutron sources and of the neutron emission along the line of sight by which the neutron source is seen by the detector to provide the direct (not collided) neutron flux and energy spectrum at the detector. 
In addition, Monte Carlo neutron transport simulations are required to estimate the scattered neutron flux and its energy spectrum given a specific neutron source and line of sight. The resulting neutron energy spectrum at the detector will contains all these spectral components at different levels. 

An example of the different components of the neutron spectrum at a detector location in a DD plasma with neutral beam heating is shown in panel (a) of figure \ref{fig:FM} which can be thought of as the neutron emissivity multiplied by the detector area and the solid angle.
The response function for each incident neutron energy is then calculated resulting in the response function matrix in which, as shown in panel (b), the columns contain the light output response function at for a particular neutron incident energy.
The response functions for incident neutron with energies 2.45 and 4.00 MeV are shown in panels (c) and (d): each response function is then folded with the energy resolution function.  
The neutron energy spectrum can then be thought as a set of weights that multiply the response function matrix for each incident neutron energy resulting in the weighted response function matrix shown in panel (e). 
Adding together the contribution from all neutron energies present in the neutron energy spectrum is then equivalent to the column-wise sum of the elements of the weighted response function matrix: the result in shown in panel (f) together with experimentally measured light output pulse height spectrum integrated over a given time interval and multiplied by the detector efficiency.
The comparison shown in panel (f) highlights one final aspect of the forward modelling, that is, the conversion of the matrix of normalized response functions multiplied by the absolute neutron energy spectrum into absolute counts. The geometry of the line of sight is already included in the neutron energy spectrum shown in panel (a) so this is achieved by multiplying the neutron energy spectrum by the integration time used to calculate the measured pulse height spectrum in panel (f) and by the detector efficiency.

In this case, the agreement is not perfect. Several reasons could be responsible for this. One possibility is that the original neutron spectrum might not have been properly estimated because one or more plasma parameters that are used in the modelling codes are measured or estimated incorrectly. Among such quantities are the fuel ion and the fast ion (from the additional heating) distribution functions in space and energy, the plasma temperature and density. If one assumes that the detector response function is well known, then such discrepancy can be used to improve the models used for the description of the plasma. 
This method of proceeding in the modelling of the measured light pulse height spectra is therefore called ``forward modelling'' since, as the name implies, it starts from the source of the neutrons and then propagates the resulting neutron field taking into account all the aspects affecting the neutron transport from the source to the detector (including realistic geometries and material compositions of the fusion device and of the diagnostic) and then converts the neutron field at the detector into the light output pulse height spectrum via the response function matrix.
It is clear therefore how crucial it is that the liquid scintillator response function is determined as accurately as possible if information on the plasma itself is required from the neutron measurements from liquid scintillators.

\section{Summary}
\label{sec:summary}
The interpretation of the neutron response function of liquid scintillators can be rather subtle as discussed in this tutorial even in the simplest case imaginable of a mono-energetic neutron source. This case, which can be approximately obtained in neutron calibration facilities, however is very useful to understand the different processes that contributes to the neutron light response function which can then easily generalized to the case of fusion neutron sources.   
Response functions can be easily obtained with Monte Carlo codes as discussed here but their use and the interpretation of the calculated results can be quite complex and as shown in this tutorial not strictly necessary for a qualitative understanding of the response function. Monte Carlo codes are however indispensable whenever a quantitative comparison with experimental measurements is crucial.
This is particularly true in the case where one wants to infer absolute quantities from measurements based on the response of liquid scintillators from fusion plasmas such for example the absolute neutron yield (which requires an accurate determination of the detector efficiency) or the underlying fuel ions distribution functions from the measured neutron energy spectrum.
In such cases, forward modelling is a robust although quite complex method that enables an understanding of the underlying plasma properties and therefore can provide useful feedback information on the theory and tools used to model the physics of fusion plasmas. This, in turns, requires a very good overall characterization of the response function of liquid scintillators. It is hoped that this tutorial provides an useful introduction to the understanding of the response functions obtained by full Monte Carlo codes.

\section*{Acknowledgements}
The author wish to thanks S. Conroy and A. Hjalmarsson at the Department of Physic and Astronomy at Uppsala University for useful discussions.



\end{document}